%% file: paper.tex
\documentclass[conference,dvipsnames]{IEEEtran}

\usepackage{cite}
\usepackage{amsmath,amssymb,amsfonts}
\usepackage{bbm}
\usepackage{algorithmic}
\usepackage{booktabs}
\usepackage{graphicx}
\usepackage{textcomp}
\usepackage[frozencache,cachedir=minted-cache]{minted}
\usepackage{xcolor}
\usepackage{url}
\usepackage{hyperref}
\usepackage[most]{tcolorbox}
\usepackage{cleveref}
\usepackage{subcaption}
\usepackage{annotate-equations}

\def\BibTeX{{\rm B\kern-.05em{\sc i\kern-.025em b}\kern-.08em
    T\kern-.1667em\lower.7ex\hbox{E}\kern-.125emX}}

\usepackage{bm}
\usepackage{booktabs}
\usepackage{todonotes}
\usepackage{listings}

\newcommand{\resulthighlight}[1]{\textbf{#1}}

\newcommand{\modelbase}{HPC-Coder-V2}
\newcommand{\datasetname}{\textsc{HPC-Instruct}}

\newcommand{\rqone}{How does the choice of fine-tuning base model and the use of instruction masking impact the performance of a code LLM on parallel code generation?}
\newcommand{\rqtwo}{How does the amount of fine-tuning data for a particular parallel execution model affect the performance of a code LLM
on that model?}
\newcommand{\rqthree}{How does the quality of parallel code fine-tuning data impact the performance of a code LLM on parallel code generation?}
\newcommand{\rqfour}{How does model size impact the ability of a code LLM to learn from distilled synthetic data?}

\definecolor{belizehole}{HTML}{2980b9}
\definecolor{LightGray}{gray}{0.9}
\definecolor{LightBlue}{HTML}{f0f8ff}
\newtcolorbox{RQcallout}[1][]{%
  colback=black!5,
  colframe=black!5,
  notitle,
  sharp corners,
  borderline west={2pt}{0pt}{belizehole!80!black},
  enhanced,
  breakable,
}

\begin{document}

\title{\modelbase{}: Studying Code LLMs Across Low-Resource Parallel Languages} %

\author{
    \IEEEauthorblockN{Aman Chaturvedi\IEEEauthorrefmark{1}\thanks{The first two authors contributed equally.}, Daniel Nichols\IEEEauthorrefmark{1}, Siddharth Singh, Abhinav Bhatele}
    \IEEEauthorblockA{~\\
        \textit{Department of Computer Science, University of Maryland, College Park, MD, USA} \\
        Email: \{achaturv, dnicho, ssingh37\}@umd.edu, bhatele@cs.umd.edu
    }
}

\maketitle

\begingroup\renewcommand\thefootnote{\IEEEauthorrefmark{1}}
\footnotetext{These authors contributed equally to this work.}
\endgroup

\begin{abstract}
\input{abstract}
\end{abstract}

\begin{IEEEkeywords}
    Large Language Models, Code Generation, HPC
\end{IEEEkeywords}

\section{Introduction}
\label{sec:introduction}
\input{intro}

\section{Background}
\label{sec:background}
\input{background}

\section{Our Approach to Improving Code LLMs for Parallel Languages}
\label{sec:overview}
\input{overview}

\section{Studies Exploring the Creation of Improved HPC Code LLMs}
\label{sec:experiments}
\input{experiments}

\section{Evaluation of Code LLMs for Parallel Code Generation}
\label{sec:evaluation}
\input{evaluation}

\section{Results of Ablation Studies}
\label{sec:results}
\input{ablation-results}

\section{\modelbase{}: An Improved Code LLM for Parallel Code Generation}
\label{sec:model}
\input{model-results}

\section{Related Work}
\label{sec:related-work}
\input{related-work}

\section{Conclusion}
\label{sec:conclusion}
\input{conclusion}

\IEEEtriggeratref{19}
\bibliographystyle{IEEEtran}
\bibliography{bib/pssg,bib/cite}

\clearpage
\appendix
\section{Appendix}
\input{appendix}

\end{document}

%% file: abstract.tex
Large Language Model (LLM) based coding tools have been tremendously successful
as software development assistants, yet they are often designed for general
purpose programming tasks and perform poorly for more specialized domains such
as high performance computing. Creating specialized models and tools for these
domains is crucial towards gaining the benefits of LLMs in areas such as HPC. While
previous work has explored HPC-specific models, LLMs still struggle to generate
parallel code and it is not at all clear what hurdles are still holding back
these LLMs and what must be done to overcome them. In this work, we conduct an
in-depth study along the many axes of fine-tuning a specialized HPC LLM in order
to better understand the challenges. Based on our findings we fine-tune and
evaluate a specialized HPC LLM that is shown to be the best performing
open-source code LLM for parallel code generation to date. 

%% file: intro.tex
Large language models (LLMs) have been a transformational technology in aiding
software development. Their ability to automate coding tasks and connect natural
language descriptions to code has improved developer productivity and enabled
developers to more rapidly move from concept to implementation. As of 2023 over
92\% of surveyed developers use AI in some form to aid their development
process~\cite{github-ai-survey}. Beyond general development assistance these
tools have the potential to enhance developer capabilities on more complex
programming tasks such as writing parallel code. Writing correct, parallel code
is an important problem facing modern developers and is already difficult for
humans. Using LLMs to improve the quality and quantity of parallel code is an
important step in improving the performance of modern software.

While code LLMs have shown promise in their code generation capabilities, they
still struggle with more complex programming tasks such as parallel code.
Previous work~\cite{nichols:hpdc2024} has extensively studied LLMs across
various parallel execution models and algorithms and found that LLMs are
significantly worse at generating parallel code compared to sequential code. Two
main reasons are identified for this discrepancy: the lack of parallel code data
in the pre-training data of modern LLMs and the intrinsic difficulty of parallel
code generation. Solving the latter issue is a long-term effort that will
require the development of more sophisticated AI models that can plan and reason
through complex problems. However, the former issue of obtaining high-quality
parallel code data at scale and effectively learning from that data is a much
more tractable problem to tackle with current language modeling capabilities.

Creating HPC and parallel capable LLMs offers a great number of benefits to the
HPC community. They will drastically improve the productivity of scientific
developers and, in turn, the speed at which scientific discoveries are made. The
process of designing these HPC capable LLMs will involve the creation of large
HPC datasets and studies into modeling that data. Building out a large corpus of
HPC data and understanding how to best learn from and model that data will be
critical to developing future HPC AI developer tools. Furthermore, as the field
of AI and code LLMs continues to progress it is important that the HPC community
understands and addresses the unique challenges associated with HPC code
generation.

Gathering parallel code data at scale and effectively learning from it is
difficult. The data samples are already underrepresented in large code datasets
and simply collecting more is often not enough; high-quality parallel code data
is needed to train models effectively. This is evinced by the results of the
StarCoder2 project which trained code LLMs on The Stack v2 dataset that contains
nearly all permissively licensed code and code related data
online~\cite{lozhkov2024starcoder}. Despite the impressive data collection
efforts, the StarCoder2 models perform similar or worse than comparable models
trained on less data. This suggests that we cannot keep improving model
performance by collecting more data, but rather we need to collect better data.
Furthermore, it is not well understood what makes data ``better'' for training
code LLMs.

In this paper we address the lack of high-quality parallel code data by creating
a large synthetic code dataset, \datasetname{}, using our proposed methodology
to map existing parallel code samples to high-quality instruct-answer pairs. We
then fine-tune code LLMs on this dataset and evaluate them against other code
LLMs on ParEval~\cite{nichols:hpdc2024}, a state-of-the-art parallel code
generation benchmark. We find that our fine-tuned model, \modelbase{}, is the
best performing open-source code LLM for parallel code generation and performs
near GPT-4 level. We conduct an in-depth study to better understand how data
representation and training parameters impact the models ability to learn how to
model parallel code. These insights will be critical for future efforts
developing the next generation of HPC AI developer tools.

In this paper we make the following important contributions.

\begin{itemize}
    \item We collect a large synthetic dataset of high quality parallel code 
        instruction data, \datasetname{}.%
    \item We fine-tune a code LLM, \modelbase{}, that is most capable open-source code LLM
        for parallel code generation.%
    \item We conduct an in-depth study along the data and fine-tuning parameters to understand
        how to best fine-tune code LLMs for parallel code generation.
\end{itemize}

Furthermore, we answer the following research questions:

\begin{itemize}
    \item[\textbf{RQ1}] \textit{\rqone{}}
    \item[\textbf{RQ2}] \textit{\rqtwo{}}
    \item[\textbf{RQ3}] \textit{\rqthree{}}
    \item[\textbf{RQ4}] \textit{\rqfour{}}
\end{itemize}

%% file: background.tex
In this section we provide background on the use of LLMs for code, LLM
distillation, and fine-tuning instruction LLMs.

\subsection{LLMs for Code}\label{sec:bg:llm-for-code}

LLMs, based on the Transformer architecture~\cite{vaswani2017attention}, have
proven to be capable of modeling text data, such as natural language and code.
Most often used for generative tasks, they can be employed in a variety of
software development tasks, such as code completion, summarization, and
translation. Building off of their success in these tasks, they are continually
being integrated into software development tools and workflows.

Code LLMs are very similar to natural language LLMs, but are generally
pre-trained, fine-tuned, and/or prompted with distinct code-specific data. For
example, popular open-source models like StarCoder2~\cite{lozhkov2024starcoder}
is pre-trained on The Stack v2 dataset~\cite{lozhkov2024starcoder}, which is a
large dataset of mostly code text. Other popular models like
CodeLlama~\cite{roziere2023code} use existing LLMs that are pre-trained on
natural language data and then fine-tuned on code-specific data. Popular code
tools like GitHub Copilot simply call existing frontier LLMs like
GPT-4o~\cite{openai2024gpt4ocard_short_author} with heavily engineered prompts
to generate code.

\subsection{LLM Knowledge Distillation}\label{sec:bg:llm-distillation}

Large frontier LLMs, like GPT-4o~\cite{openai2024gpt4ocard_short_author},
generally give the best responses across a large variety of tasks, however, they
are computationally and financially expensive to run. For this reason, the
practice of knowledge
distillation~\cite{xu2024surveyknowledgedistillationlarge}, where a smaller
model is trained to be as good as a larger model for a particular sub-task, is
becoming increasingly popular. Knowledge distillation techniques generally
either employ a teacher-student model, where the teacher is the larger model and
the student is the smaller model, or a model compression technique, where the
larger model is   
compressed into a smaller model. This work focuses on a simple form of
teacher-student knowledge distillation, where the large model is used to
generate lots of high quality synthetic data samples that are then used to train
a smaller model.

\subsection{Fine-tuning Instruction LLMs}\label{sec:bg:instruction-llms}

Instruction LLMs are a specialized form of LLMs that are fine-tuned to receive a
natural language instruction from the user and generate a response. They behave
like a chatbot, but do not necessarily handle multi-turn dialog. These are
usually created by first selecting a general LLM that was pre-trained on a
corpus of general text data and then fine-tuning on a corpus of dialog data.
This is accomplished by showing the model samples in the format ``Instruct:
\{instruction\} Response: \{response\}'' and then training the model to generate
the response. This is generally very effective at getting LLMs to follow user
prompts and most models available today have an instruction variant available. 

Generally, instruction LLMs are fine-tuned using \textbf{\emph{
instruction-masking}}. When fine-tuning with instruction masking, the gradient
values corresponding to the instruction tokens are masked to zero to prevent the
model from learning to generate the instruction tokens. Instead, weights are
only updated based on its ability to predict missing tokens in the response.
Conceptually, this is done since there may be bad text in the instruction that
we do not want the model to learn to generate. For example, the user instructs
the model to fix their buggy code. In this case the instruction will contain bad
code, which we do not want the model to learn to generate. Instead, we want the
model to learn to generate the fixed code in the response. While {\it
instruction-masking} is common practice and conceptually clear, there is little
literature arguing quantitatively for its effectiveness.

%% file: overview.tex
\begin{figure}[h]
    \centering
    \includegraphics[width=\linewidth]{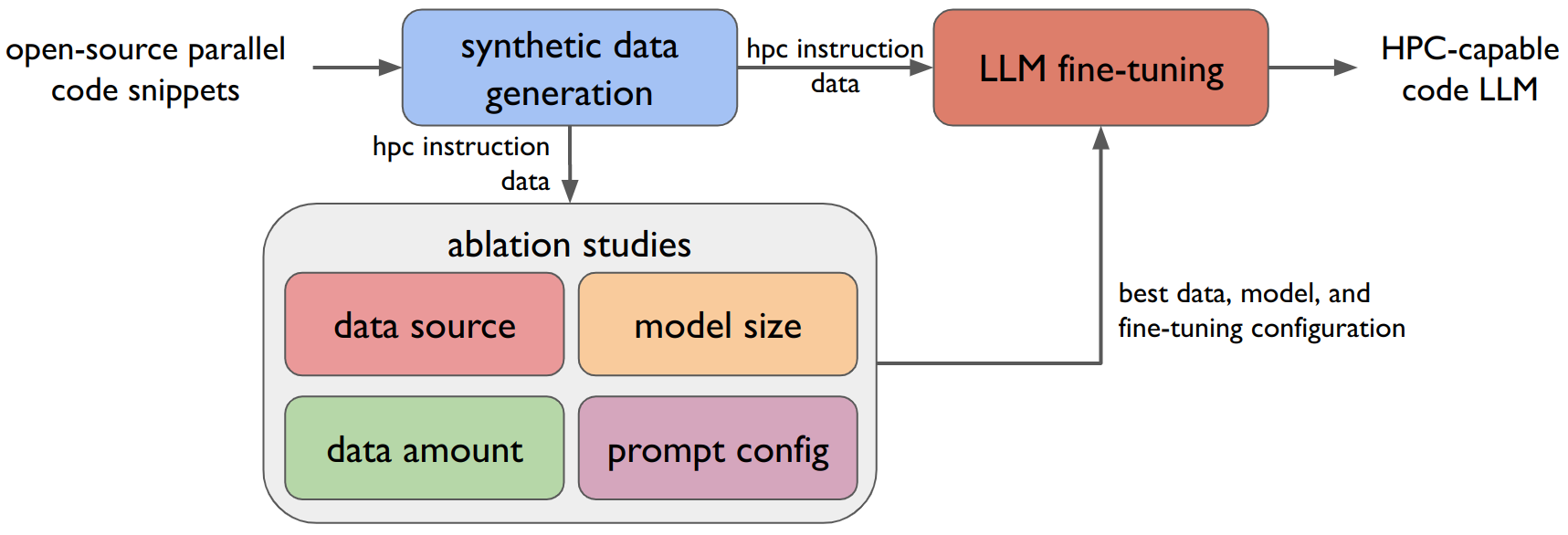}
    \caption{Overview of the methodology proposed in this paper. First, we use
        open-source parallel code snippets to generate a large synthetic
        instruction dataset of parallel code samples. We then conduct ablation
        studies to understand how data, model, and fine-tuning parameters impact
        the capability of a code LLM to write parallel code. Finally, we utilize
        the dataset and insights from the ablation studies to fine-tune a code
        LLM for parallel code generation and evaluate it against other code LLMs
        on the parallel code generation benchmark ParEval.
        \label{fig:overview}}
\end{figure}

Our approach to improving Code LLMs for parallel languages involves creating a
large synthetic code dataset, \datasetname{}, and then fine-tuning existing
pre-trained Code LLMs on this dataset.  We first present an overview of our
proposed approach (\Cref{fig:overview}) and then present details of the various
components.

We begin by generating a large scale synthetic dataset of code
samples using open-source parallel code snippets and state-of-the-art LLMs. This
dataset is comprised of roughly 120k parallel code instruction-response pairs
where the instruction is a natural language problem description and the response
is the code that solves the problem. The construction of this dataset is
inspired by previous work~\cite{wei2023magicoder} that demonstrated the success of
fine-tuning smaller code LLMs on synthetic data generated from larger foundation
models.

Using the HPC instruction dataset, we then conduct an in-depth study along the
axes of code model fine-tuning to better understand how data representation and
quality, model size, and prompt construction impact the ability of a code LLM to
learn how to generate parallel code. During these studies we evaluate each of
the fine-tuned models against the ParEval~\cite{nichols:hpdc2024} benchmark to
understand their performance on real parallel code generation tasks. These
studies yield critical insights into best practices for fine-tuning HPC
code LLMs.

Finally, with the full HPC instruction dataset and insights from the ablation
studies, we fine-tune three state-of-the-art HPC capable code LLMs. These are
evaluated against the ParEval benchmark and compared to other state-of-the-art
LLMs for their ability to generate parallel code.

\subsection{Ameliorating the Data Problem with Synthetic Data}
\label{sec:data-collection}
\input{data-collection}

\subsection{Selecting a Pre-trained Model}\label{sec:fine-tuning:base-model}

Before fine-tuning, we need to first select a pre-trained model to fine-tune. When
fine-tuning smaller open-source models, choosing a model already trained
for code tasks tends to yield better results~\cite{bigcode_leaderboard}. Based
on this and the successful results of previous code LLM fine-tuning
studies~\cite{wei2023magicoder} we select the
DeepSeek-Coder~\cite{guo2024deepseekcoder, deepseek-coder-v2} family of models
for fine-tuning. In particular, we fine-tuned the 1.3b,
6.7b~\cite{guo2024deepseekcoder}, and 16b~\cite{deepseek-coder-v2} parameter
models. These models are state-of-the-art in code modeling and outperform other
LLMs on many coding benchmarks~\cite{bigcode_leaderboard, guo2024deepseekcoder,
deepseek-coder-v2}. They are trained on a dataset of 87\% code and 13\% natural
language with a 16k context window.  The 1.3b and 6.7b are based on the
llama~\cite{touvron2023llama} model architecture, while the 16b is a custom
mixture-of-experts (MOE)~\cite{fb-moe} architecture. The MOE architecture
enables the 16b model to scale to larger sizes while maintaining runtime
performance.

\subsection{Fine-Tuning on Synthetic HPC Code Data}
\label{sec:fine-tuning}
\input{fine-tuning}

%% file: data-collection.tex
Before we can fine-tune HPC LLMs, we need to collect a large dataset of HPC
relevant code and dialog. While large datasets of open-source code
exist~\cite{lozhkov2024starcoder}, previous work has shown that generating
structured synthetic data with state-of-the-art LLMs can yield data much more
effective for fine-tuning specialized code LLMs~\cite{wei2023magicoder}. This
section details our approach to collecting large-scale synthetic data for HPC
based on this insight.

\begin{figure*}[h]
    \centering
    \includegraphics[width=\linewidth]{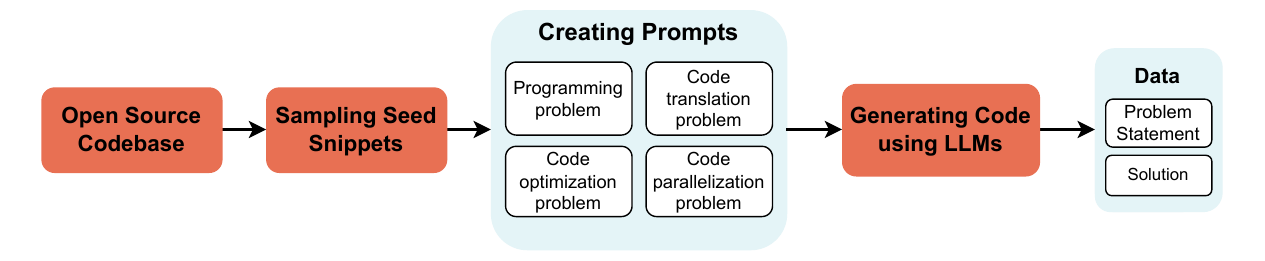}
    \caption{Synthetic data generation process. We collect seed snippets from
    open source codebases and combine them with multiple prompt templates to
    create data generation prompts for an LLM. These prompts are then used to
    generate problem-solution pairs with an LLM. \label{fig:synthetic-data}}
\end{figure*}

\begin{figure}
    \centering
    \includegraphics[width=\linewidth]{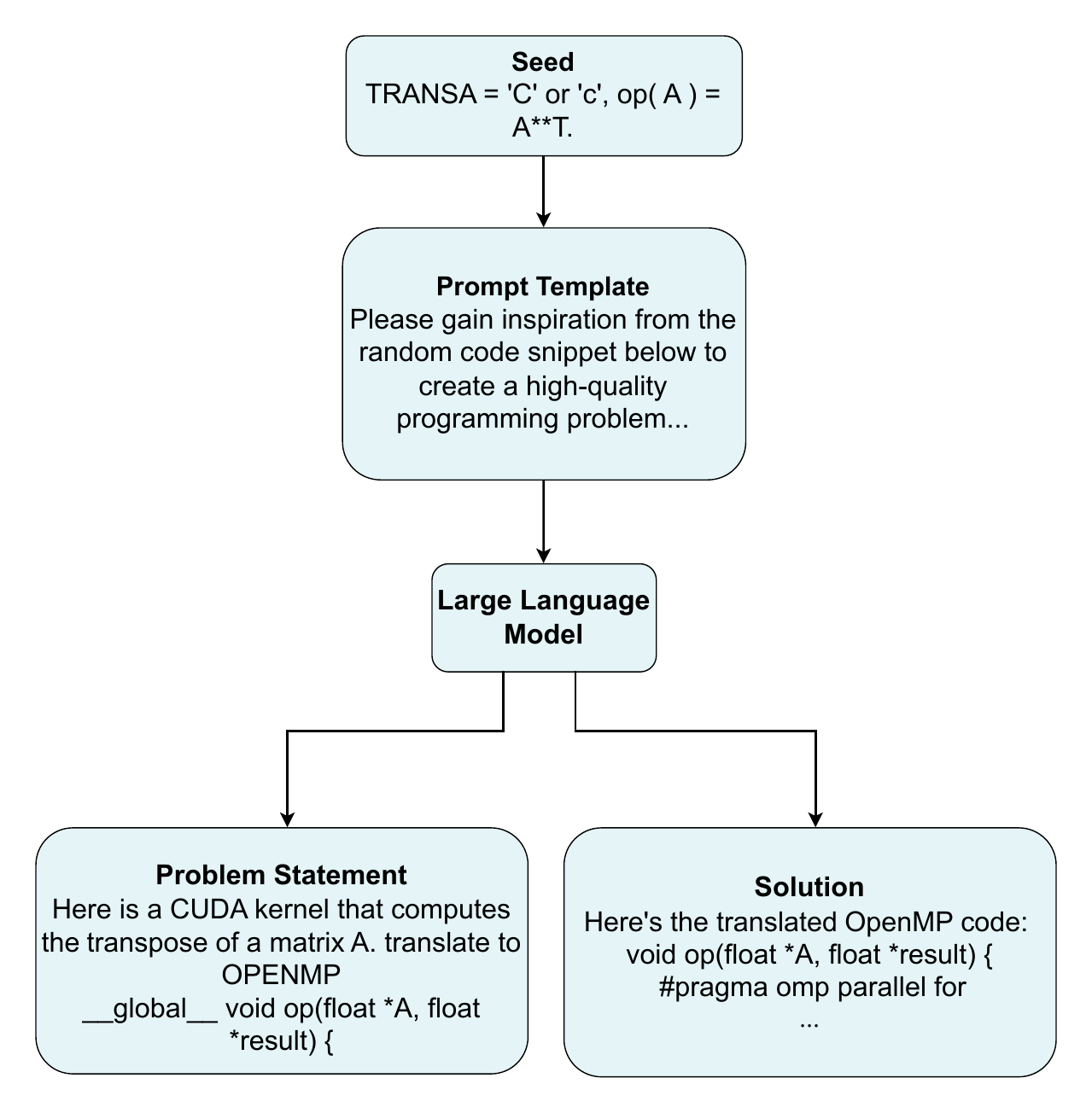}
    \caption{Example synthetic data generation output. Here, a random seed
    snippet is used alongside the translation prompt template and fed into the
    LLM. The resulting synthetic sample from the LLM is a problem of translating
    some code to OpenMP and the corresponding solution.
    \label{fig:example}}
\end{figure}

While state-of-the-art commercial LLMs like GPT-4o can generate high-quality
instruction samples, they tend to generate very repetitive samples. To address
this, we adapt the use of seed code snippets from~\cite{wei2023magicoder} to get
diverse outputs from the LLM. We gather a diverse set of seed snippets from
open-source codebases in The Stack V2~\cite{lozhkov2024starcoder}, focusing on
code in HPC languages (C, Fortran, etc.) and using HPC libraries (MPI, OpenMP,
etc.). In total we collect 125k seed snippets including 25,000 samples in
Python, C, FORTRAN, and C++, 15,000 samples in CUDA, and 5,000 samples in Chapel
and OpenCL. When asked to generate a data sample, the LLM is asked to be
inspired by the seed snippet, yielding more diverse and creative outputs. We
obtain further variety in the generated data by generating multiple sample
types: programming, translation, optimization and parallelization problems. This
process is visualized in~\Cref{fig:synthetic-data}. An example programming
template response can be seen in~\Cref{fig:example}, illustrating the workflow
from seed snippet selection to the final dataset sample.

\vspace{0.2em}
\noindent\textit{Programming Prompts:} In this template, the LLM is tasked with
generating a parallel programming problem and a corresponding solution.

\vspace{0.2em}
\noindent\textit{Translation Prompts:} The translation templates
directs the LLM to create a problem focused on converting code from one parallel
programming language to another. For example, the model might be prompted to
translate a CUDA-based implementation into OpenMP or OpenMP to MPI. 

\vspace{0.2em}
\noindent\textit{Optimization Prompts:} For these prompts, we ask the LLM to
generate an optimization problem and a corresponding solution.

\vspace{0.2em}
\noindent\textit{Parallelization Prompts:} The parallelization template
asks the LLM to parallelize a given code snippet, transforming it from a
sequential implementation to an efficient parallel version. 

Using the 125k formatted prompts we generate synthetic data samples with four
state-of-the-art LLMs: Gemini-Pro, DBRX, Llama-3-70B, and Mixtral-8x7B. The
resulting dataset, named \datasetname{}, comprises over 122k synthetic data
samples (some outputs were not parsable and discarded). We use several LLMs to
gather a variety of samples, further ensuring data diversity. It also enables us
to study the impact of data quality along the axis of source generation model.
An example data sample from \datasetname{} is shown
in~\Cref{lst:appendix:example-problem}.

%% file: fine-tuning.tex
We fine-tune each of the models on the \datasetname{},
Magicoder-OSS-Instruct-75K~\cite{wei2023magicoder}, and
Evol-Instruct-Code-80k-v1~\cite{luo2023wizardcoder} datasets. The latter two
datasets are state-of-the-art synthetic and semi-synthetic code instruction
datasets. We include these since, although they are not HPC specific, they can
still improve the model's generalization capabilities. In total the fine-tuning
dataset has 277k samples.

We use the AxoNN~\cite{singh:ipdps2023} framework to fine-tune the models. This
is a parallel deep learning framework wrapped around
PyTorch~\cite{paszke2019pytorch}. It handles automatically parallelizing the
model across GPUs and allows us to fine-tune the models that do not fit in
memory on a single node. The 6.7b and 16b models are fine-tuned on four nodes
each with four 80GB A100 GPUs, while the 1.3b model is fine-tuned on two A100
GPUs. The total fine-tuning times range between 3 and 20 hours.

We fine-tune the 1.3b and 6.7b models in bfloat16 precision with a batch size of
128 and a sequence length of 8192 for two epochs. The 16b model is fine-tuned
with a batch size of 1024 for one epoch. Furthermore, we employ the AdamW
optimizer~\cite{adamw} to update the model weights based on the fine-tuning
loss. This training setup and hyperparameters are selected based on those used
in related literature to fine-tune code LLMs. Cursory experiments showed that
these hyperparameters work well for our fine-tuning task, however, it is
possible that an exhaustive search could yield better results. Performance
hyperparameters, like batch size, are selected based on the model size,
available GPU memory, and desired performance. The context window length is
lowered from 16k to 8k from the base models, since none of the data samples in
the dataset exceed 8k tokens and this saves memory and performance during
fine-tuning.

%% file: experiments.tex
We now have a large dataset of synthetic instruction HPC data, \datasetname{},
and the setup to fine-tune code LLMs on it. However, there are many unknowns
regarding how to best fine-tune these models such as how to format prompts, how
much and what quality of data to use, what size of model to use, etc. In this
section, we present a series of ablation studies along different axes of model
fine-tuning to better understand how each contributes to the ability of a
fine-tuned code LLM to generate parallel code.

\subsection{Choice of Base Model and Instruction Masking}

In this experiment, we fine-tune the Deepseek-Coder 1.3B and 6.7B base and
instruct models with and without instruction masking on the \datasetname{},
Magicoder-OSS-Instruct-75K, and Evol-Instruct-Code-80k-v1 datasets. In total
there are eight models fine-tuned: \{1.3B, 6.7B\} $\times$ \{base, instruct\}
$\times$ \{masked, unmasked\}. We omit the 16B model from this experiment due to
its high computational cost for fine-tuning. The goal of this experiment is to
better understand the impact of the choice of base model and instruction
masking. We choose to study the impact of base versus instruct models as it is
unclear from related literature which model type is better for fine-tuning on
specific tasks. Generally, most users interact with instruct models as they are
able to follow instructions and better engage in dialog-like interactions. For
this reason, most open-source models have instruct models available that have
been fine-tuned from a base model. When fine-tuning a new instruct model, on one
hand, it may be better to reap the benefits of the existing fine-tuned model and
start from there. On the other hand, it may be better to start from scratch with
a base model, since they will be more general and easier to fine-tune.

We also study the impact of instruction masking
(see~\Cref{sec:bg:instruction-llms}). Instruction masking is usually employed to
prevent the model from learning bad patterns that may be present in the user
instruction. We only want to learn from the responses. While intuitive, we are
actually learning from less information when we mask the instruction and it is
unclear if this trade-off between learning from less information and learning
from less noise is worth it.

\subsection{Studying the Impact of the Amount and Quality of Parallel Code Data}

Even with an ideal base model and prompting strategy it is still difficult to
fine-tune a good model without the right amount and quality of data. To further
explore this, we conduct two experiments: one to study the impact of the amount
of data from individual parallel models and another to study the impact of the
quality of data.

For the first experiment, we create several versions of the \datasetname{} each
with varying amounts of MPI code samples: 0k, 2k, 4k, 6k, 8k, 10k, and 12k. We
leave the other data in the dataset unchanged and just vary the amount of MPI
data. MPI samples are identified by the presence of certain substrings like
``mpi.h'' or ``MPI\_Init'' in the code. These datasets are used to fine-tune the
1.3B and 6.7B models resulting in 14 total models. The purpose of this study is
to shed light on how the amount of data from a specific parallel model affects
the final performance of the LLM on that parallel model. Does performance keep
increasing with more data or does it plateau at some point? This is important as
it informs how we collect future data for fine-tuning. We select MPI for this
study as LLMs consistently perform worse at generating MPI code than any other
parallel programming model~\cite{nichols:hpdc2024} and, therefore, it is
desirable to improve their ability to generate MPI code.

Tangetially, we also study the impact of the quality of data on the fine-tuned
models. As LLMs are increasingly getting more dependent on synthetic data for
training, it is also getting extremely important to validate the quality of the
synthetic data being produced to see its effect on model performance. We
hypothesize that there is a trade-off between the amount of data and the quality
of data, where eventually more data stops improving performance and quality
becomes more important. Understanding this trade-off is particularly vital for
synthetic data where we are expending compute to create the data; we need to
know whether compute time is better spent on more data or better data.

Directly studying data quality is difficult as it is hard to quantify and the
scale of data is too large for qualitative analysis. In order to overcome this
we instead use the base model used for generating the synthetic data as a proxy
for differences in data quality. We presume that the diffent models generate
data of different quality. This will not allow us to infer what makes the data
better or worse, but it will allow us to see if quality impacts the ability of
the fine-tuned model to generate parallel code. To conduct this experiment we
fine-tune the 1.3B and 6.7B models on the \datasetname{} dataset generated from
four different LLMs: Gemini-Pro, DBRX, Llama-3-70B, and Mixtral-8x7B. We also
fine-tune both models on all of the data together. This results in ten total
models that we can compare to see if the quality of the data impacts the final
performance of the fine-tuned model.

\subsection{Studying the Impact of Model Size}

Finally, we aim to study how model size impacts the final performance of a
fine-tuned model. While larger models tend to be better at most tasks, there is
a trade-off where the time and resources necessary to run a larger model may not
be worth the marginal increase in performance. For example, if a 7B parameter
model is able to generate code for a particular niche task {\it nearly} as well
as a 70B parameter model, then it is likely much more practical for a user to
simply use the 7B model. It will run quickly on a consumer laptop whereas the
70B model will require specialized hosting or multiple GPUs. To study the impact
of model size, we fine-tune the 1.3B, 6.7B, and 16B models on the \datasetname{}
dataset. This will allow us to compare the performance of the models across
different sizes and see if the larger models are worth the extra resources.

%% file: evaluation.tex
With many different versions of the models fine-tuned, we now need a way to
evaluate their efficacy for parallel code generation and compare them. This will
allow us to understand the impact of different fine-tuning and data setups on
the final model performance. In this section we detail the benchmarks and
metrics used to compare models for parallel code generation.

\subsection{Benchmark Used}

When evaluating LLMs for code generation it is imperative to evaluate them on
code correctness. To do this for parallel code generation we use the
state-of-the-art benchmark ParEval~\cite{nichols:hpdc2024}. ParEval has 420
coding problems that it uses to test an LLM's parallel code generation
capabilities. These problems range across 12 different problem types: sort,
scan, dense linear algebra, sparse linear algebra, search, reduce, histogram,
stencil, graph, geometry, fourier transform and transform help us show the
diversity on which the model has been tested on. For each of the problem types
there are problems across seven different execution models: mpi, mpi+omp, cuda,
kokkos, serial, hip, omp. ParEval provides drivers to run and unit test the
generated code for correctness. Furthermore, the results can be analyzed along
the many different axes of the problem types and execution models.

We also compared our model's memory requirements and throughput with other
models to better understand the trade-offs between model size, performance and
accuracy. These numbers are recorded on the ParEval benchmark when generating
outputs using an H100 and a batch size of one. These results are important to
users who may be constrained by hardware with limited memory or speed.

\subsection{Metrics for Comparison}

Since LLMs are probabilistic and may output different results for the same
problem it is generally best to evaluate them in a probabilistic manner. For
code LLMs most papers have adopted the pass@$k$ metric to do
this~\cite{codex-copilot-short-author}. This metric quantifies the probability
that an LLM can generate at least one correct solution within $k$ attempts.
Since we cannot calculate this probability directly we need to estimate it. To
do this for one prompt, N samples are generated where N is much greater than k,
which are then evaluated on code correctness and used to estimate pass@k.
Choosing N to be much greater than k ensures that we can compute a statistically
significant estimate of pass@k. The pass@$k$ compute is shown
in~\Cref{eq:pass-k}.

\vspace{1em}
\begin{equation}\label{eq:pass-k}
    \text{pass@}k =
    \frac{1}{\lvert \eqnmarkbox[MidnightBlue]{P1}{P}\rvert}
    \sum_{p\in \eqnmarkbox[MidnightBlue]{P2}{P}}
    \left[
        1 -
        \binom{
            \eqnmarkbox[WildStrawberry]{N1}{N} -
            \eqnmarkbox[OliveGreen]{cp}{c_p}
        }{
            k
        }
        /
        \binom{
            \eqnmarkbox[WildStrawberry]{N2}{N}
        }{
            k
        }
    \right]
\end{equation}
\annotatetwo[yshift=1em]{above}{N1}{N2}{Number of samples generated per prompt}
\annotatetwo[yshift=-1em]{below}{P1}{P2}{Set of prompts}
\annotate[yshift=-2.3em]{below,right}{cp}{Number of correct\\samples for prompt $p$}
\vspace{1.5em}

To further demonstrate pass@$k$, say we want to generate a pass@1 score for a
model, it will generate \(N=10\) samples for a given prompt and out of these
\(c_p=3\) samples are correct. Using the formula, we will get a score of 0.3 so
the model has a 30 percent chance of generating the correct solution in it's
first attempt. The pass@1 metric is an important benchmark that is used to
evaluate models' usability which is why we use it to compare our model with
other models to see where it stands. In recent years, papers have resorted to
just reporting pass@$k$ for $k=1$ as LLMs have become more powerful and can
generate correct code more often. It is also a more desirable metric for the
user who wants code to be generated correctly the first time.

\subsection{Other Models Used for Evaluation}

We compare our final models with several other state-of-the-art code LLMs to
better understand their performance and our study's insights can lead to
improvements in the field. We compare our models with the following models:

\begin{itemize}
    \item \textbf{StarCoder2 (1.3B, 7B, 15B)}: LLMs pre-trained on a large
        corpus of mostly code data from The Stack V2~\cite{lozhkov2024starcoder}.
    \item \textbf{Magicoder (6.7B)}: A fine-tuning of the DeepseekCoder-6.7B
        model fine-tuned on synthetic data generated based on open-source code~\cite{wei2023magicoder}.
    \item \textbf{Phind-V2 (34B)}: A fine-tuning of the
        CodeLlama-34B~\cite{roziere2023code} model on a proprietary
        dataset~\cite{phind-codellama-34b-v2}. At the time of its release it was
        the best model on the BigCode leaderboard~\cite{bigcode_leaderboard}.
    \item \textbf{Gemini-1.5-flash}: A commercial model avaiable via API
        from Google~\cite{geminishort2023gemini}.
    \item \textbf{GPT-3.5, GPT-4}: State-of-the-art commercial LLMs from OpenAI
        only accessible via API~\cite{gpt-3,openai2023gpt4}. 
\end{itemize}

%% file: ablation-results.tex
With the different models trained across the various configurations and 
data partitions, we can now analyze each model's parallel code generation
performance to better understand the impact of different training configurations.
In this section we detail the results from each of these ablation studies and
provide insights into how to best train an HPC specialized code LLM.

\subsection{Choice of Base Model and Instruction Masking}\label{sec:results:configuration}

\begin{RQcallout}
    {\bf RQ1 }{\it \rqone{}}
\end{RQcallout}

\begin{figure}[h]
    \centering
    \includegraphics[width=\linewidth]{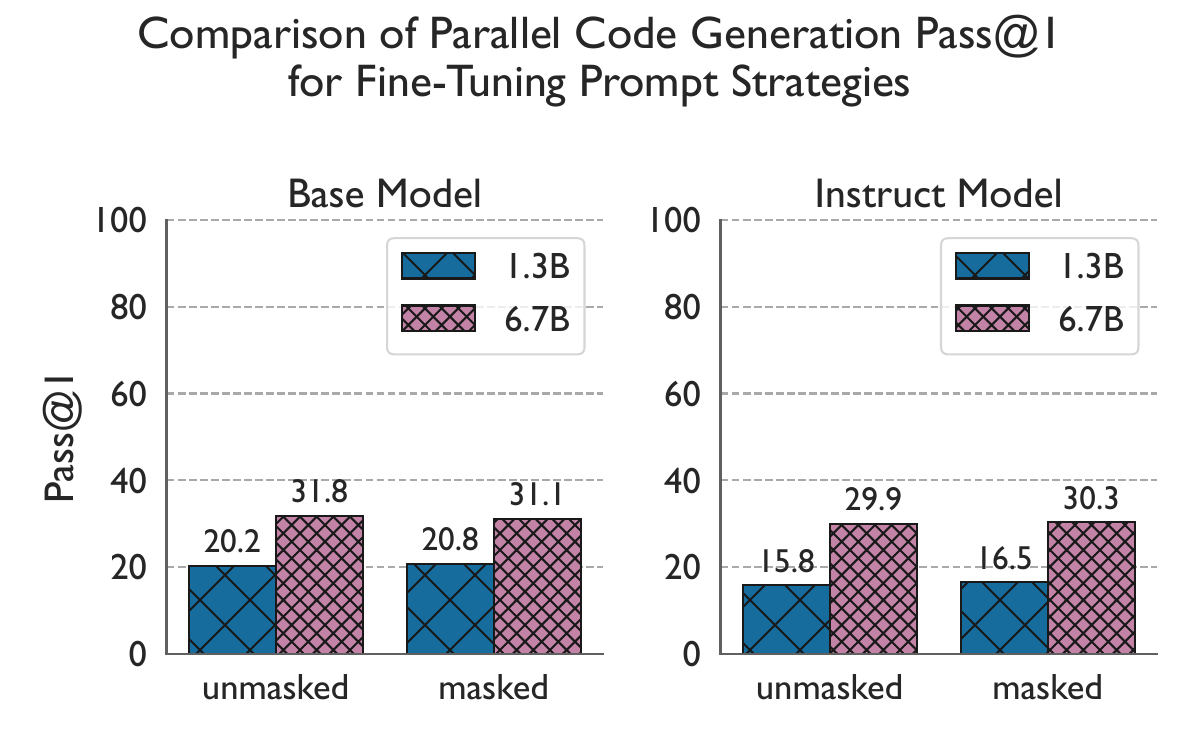}
    \caption{ParEval parallel code generation scores for various prompt formats.
    Results are shown for 8 total model configurations: \{masked, unmasked\}
    gradients $\times$ \{instruct, non-instruct\} base models $\times$ \{1.3B,
    6.7B\} model sizes. There is no correlation in parallel code generation
    performance between masked and unmasked gradients, however, fine-tuning the
    base model rather than the instruct gives much better results for both 1.3B
    and 6.7B models. \label{fig:results:prompt-format}}
\end{figure}

\Cref{fig:results:prompt-format} details the parallel code generation results on
ParEval for the masked/unmasked and instruct/non-instruct prompt formats. There
are eight models shown in the figure; they were fine-tuned on the Deepseek-Coder
base models and the Deepseek-Coder instruct models using either masked or
unmasked gradients. \resulthighlight{We observe little correlation between using
masked and unmasked gradients on the instruction prompts}. Using masked
gradients instead of unmasked provides a slight less than one percentage point
improvement for the 1.3B models. However, using masked gradients hurt
performance when fine-tuning the 6.7B model. This goes against traditional
wisdom that using masked gradients is better for fine-tuning instruction models.

Unlike for masking, there is a notable difference between fine-tuning the base
version of a model and an existing instruct variant. \resulthighlight{We observe
that fine-tuning base models, rather than instruct variants, leads to better
performance at parallel code generation.} This is true across all
configurations: 1.3B and 6.7B models, masked and unmasked gradients. The
difference is most pronounced for the 1.3B models, where fine-tuning the base
models gives a roughly 4 percentage point advantage over fine-tuning the
instruct models. While it is difficult to pinpoint the exact cause of this
difference, it is likely that the instruct models were fine-tuned to model a
less general distribution when they were first fine-tuned from the base model.
In other words, it is better to fine-tune base models and not fine-tunings of
them, since the base models are more general and can be fine-tuned to a specific
task more effectively.

\subsection{Studying the Impact of the Amount and Quality of Parallel Code Data}\label{sec:results:data-ablation}

\begin{RQcallout}
    {\bf RQ2 }{\it \rqtwo{}}
\end{RQcallout}

\begin{figure}[h]
    \centering
    \includegraphics[width=\linewidth]{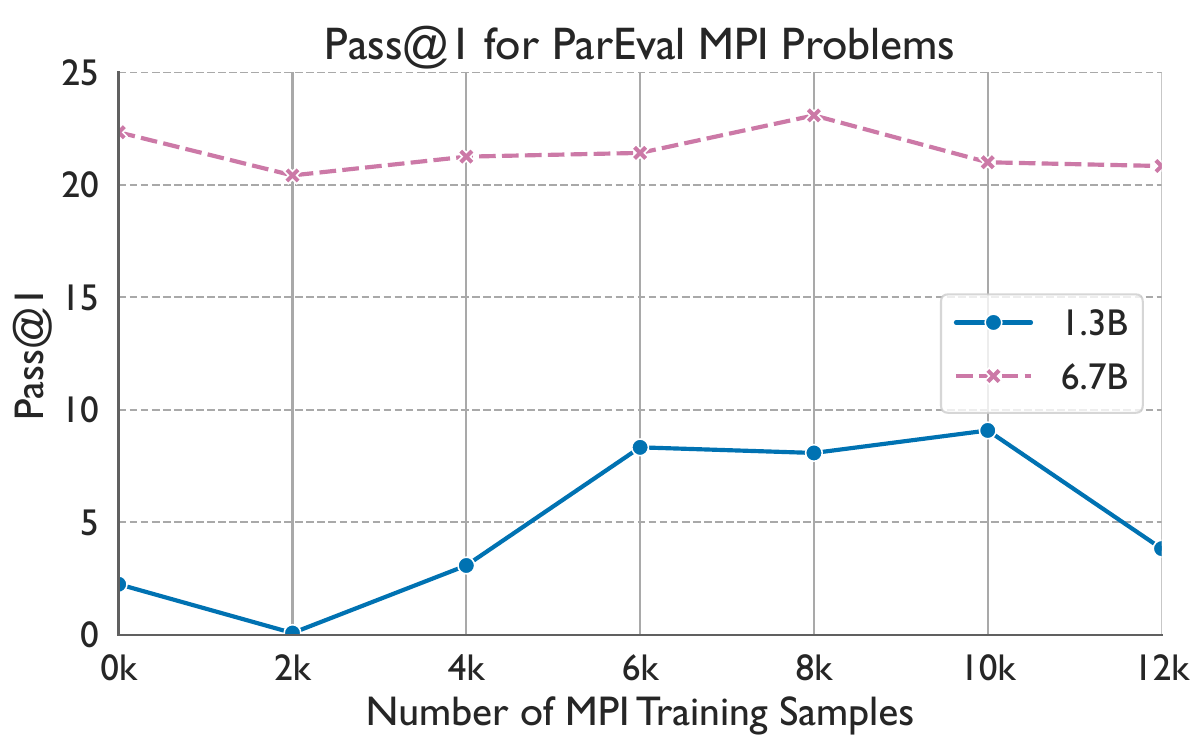}
    \caption{ParEval MPI code generation performance for increasing amounts of
        MPI fine-tuning date. As the amount of MPI fine-tuning date increases
        the smaller 1.3B model sees an increase in ability to generate MPI code
        with diminishing returns after 6k samples. The larger 6.7B model sees no
        improvement in MPI code generation performance with additional data.
        \label{fig:results:mpi-ablation}}
\end{figure}

\Cref{fig:results:mpi-ablation} presents the MPI code generation performance for
various amounts of MPI fine-tuning data. MPI is selected for this study since
LLMs consistently perform worse at generating MPI code than any other parallel
execution model~\cite{nichols:hpdc2024} and, therefore, it is desirable to
improve their ability to generate MPI code. In total there are 14 models shown
in the figure: the 1.3B and 6.7B Deepseek-Coder models each fine-tuned on
datasets with 0k, 2k, 4k, 6k, 8k, 10k, and 12k MPI samples. After running
ParEval's MPI benchmarks on these models, \resulthighlight{we observe that
increasing the amount of training data for a particular parallel execution model
can improve the performance of smaller code LLMs on that model with diminishing
returns, but has little to no effect on larger models.} 

The 1.3B models see a gradual increase in MPI code generation performance until
6k MPI samples, after which the performance plateaus and eventually decreases at
12k MPI samples. The plateau can be explained by smaller models being more
susceptible to overfitting. The 6.7B models, on the other hand, have fairly
consistent MPI code generation performance across all amounts of MPI fine-tuning
data. The model has already learned all it can from the data and adding more has
no effect on performance.

\begin{RQcallout}
    {\bf RQ3 }{\it \rqthree{}}
\end{RQcallout}

\begin{figure}[h]
    \centering
    \includegraphics[width=\linewidth]{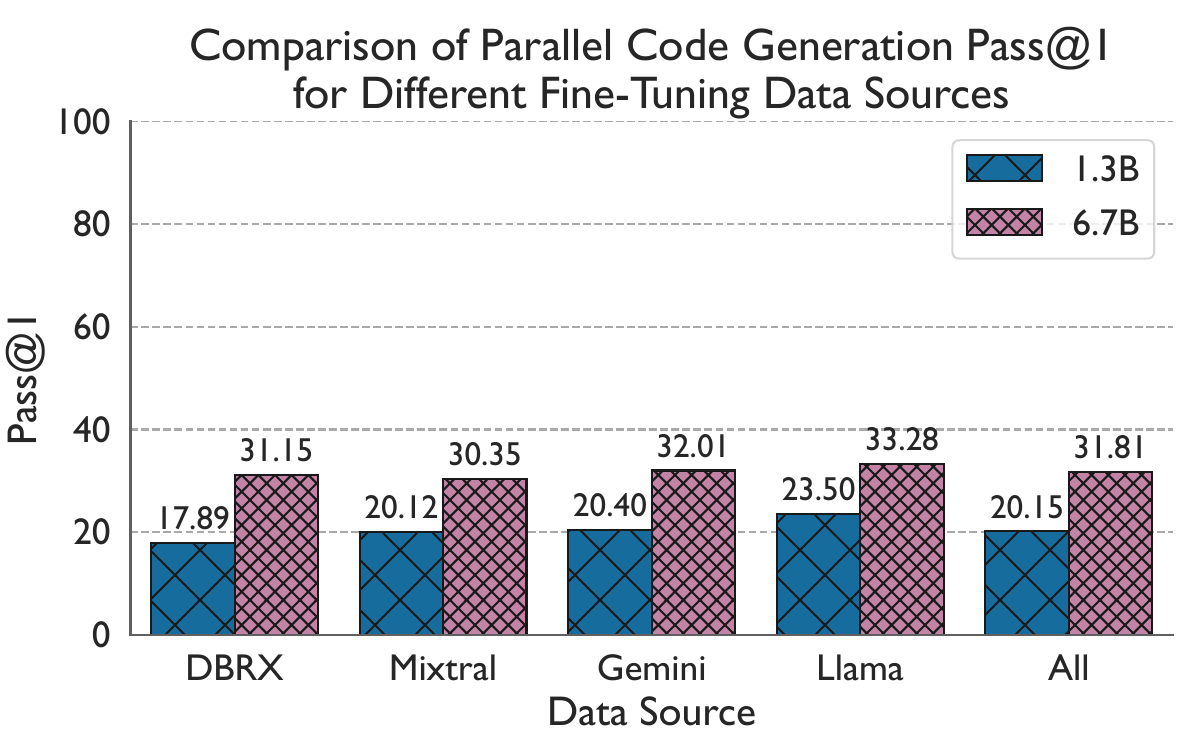}
    \caption{ParEval parallel code generation performance across different
        synthetic data sources. There is a clear difference in performance across
        data sources with Llama generated synthetic data leading to the best performing
        LLMs and DBRX leading to the worst.\label{fig:results:data-ablation}}
\end{figure}

In addition to the amount of data, the quality of the data can also impact the
ability of an LLM to learn from it. To study this, we examine the performance of
the models when fine-tuned on \datasetname{} synthetic data with different LLMs
used to generate the data. \Cref{fig:results:data-ablation} shows the ParEval
performance of each of these models. \resulthighlight{We observe that the
quality of the parallel code fine-tuning data can have a significant impact on
the performance of a code LLM on parallel code generation.} Models trained on
Llama3-70B generated data have up to six percentage points higher parallel code
generation performance than those trained on DBRX data. While it is difficult
quantify the quality of these data samples, it is clear that the quality of the
data does lead to a measurable difference in generation quality. This motivates
further investigation into what makes a training data sample high quality.

\subsection{Studying the Impact of Model Size}\label{sec:results:model-size}

\begin{RQcallout}
    {\bf RQ4 }{\it \rqfour{}}
\end{RQcallout}

\begin{figure}[h]
    \centering
    \includegraphics[width=\linewidth]{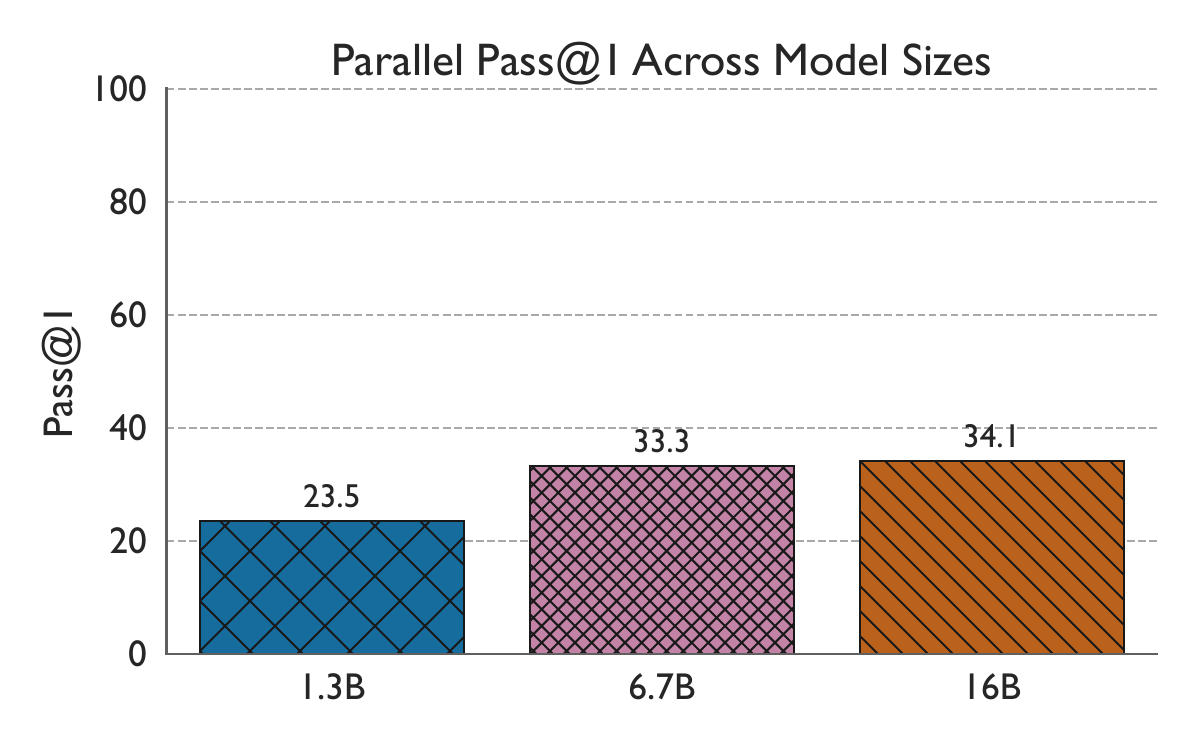}
    \caption{ParEval serial and parallel code generation performance along
        various base model sizes. There is a significant increase in performance
        from 1.3B to 6.7B, but a much smaller increase from 6.7B to 16B.
        \label{fig:results:model-size-ablation}}
\end{figure}

Finally, we investigate the impact of base model size when fine-tuning a code
LLM. This is a crucial question as larger models are considerably more expensive
to fine-tune, store, and deploy for inference. Understanding the trade-offs
between size and generative capabilities is essential for designing practical
code LLMs. \Cref{fig:results:model-size-ablation} shows the ParEval performance
of the 1.3B, 6.7B, and 16B models fine-tuned on the same \datasetname{} data.
\resulthighlight{We observe a significant increase in performance from 1.3B to
6.7B, but a much smaller increase from 6.7B to 16B.}

The diminishing return as model size increases is expected as we are using
knowledge distillation to train the models; the performance of the LLMs is
unlikely to surpass the performance of the teacher model. Based on the ParEval
results in~\cite{nichols:hpdc2024}, the 16B model is approaching the parallel
code generation performance of foundation models like GPT-3.5 and GPT-4.

%% file: model-results.tex
Using the insights from the ablation studies we train a series of models with
the best configuration to create state-of-the-art parallel code generation LLMs.
In this section we evaluate these models, \modelbase-1.3B, \modelbase-6.7B, and
\modelbase-16B, on the ParEval benchmark suite and compare their performance
with other state-of-the-art code LLMs. \Cref{sec:appendix:full-results} has the
full ParEval results for all models.

\subsection{\modelbase{} Across Problem Types and Execution Models}\label{sec:results:problem-type}

\begin{figure*}[th]
    \centering
    \includegraphics[width=\linewidth]{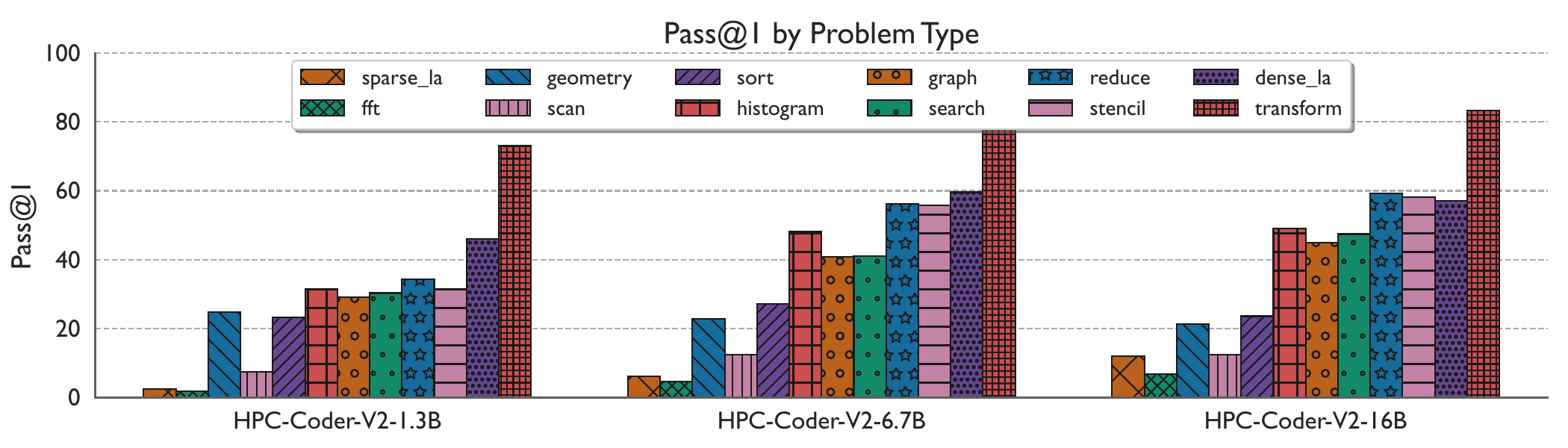}
    \caption{ParEval code generation performance by problem type. These results
        follow similar trends to those shown in~\cite{nichols:hpdc2024} except
        with higher performance across all problem types.
        \label{fig:results:problem-type}}
\end{figure*}

\Cref{fig:results:problem-type} shows the code generation performance of
\modelbase{} across the twelve problem types in the ParEval benchmark suite. We
observe similar trends to those shown in~\cite{nichols:hpdc2024} except with
higher performance across all problem types. The LLMs tend to struggle with
sparse unstructured problems, such as {\it sparse linear algebra} and {\it
geometric} problems. The models perform much better on dense, structured
problems such as {\it dense linear algebra}, {\it stencil}, and simple {\it data
transformation} problems. With the exception of {\it geometric} problems, the
models perform better as their size increases with the 16B model performing the
best across all problem types. Interestingly, the models perform worse on {\it
geometric} problems as the model size increases.

\begin{figure}[h]
    \centering
    \includegraphics[width=\linewidth]{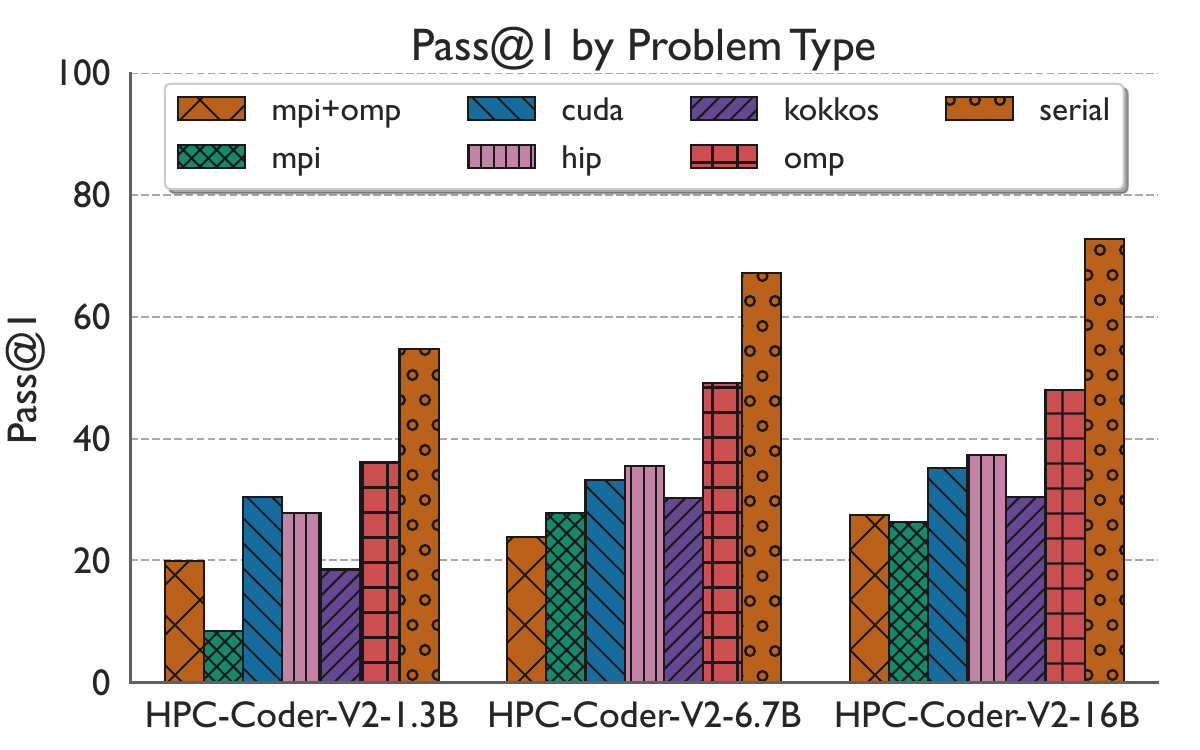}
    \caption{ParEval code generation performance by execution model. The LLMs
        perform best on serial code followed by OpenMP. The models struggle most
        with MPI code generation. \label{fig:results:execution-model}}
\end{figure}

Another axis of comparison besides problem type is the parallel execution model.
\Cref{fig:results:execution-model} shows the code generation performance of the
three LLMs across the seven execution models in ParEval. As with the problem
types we see similar trends as in~\cite{nichols:hpdc2024}. The LLMs always
perform best on {\it serial} code followed by {\it OpenMP}. This is expected as
OpenMP code is most similar to its serial counterpart. The next best performing
execution models are the GPU models, {\it CUDA} and {\it HIP}. These are
followed by {\it Kokkos} and the MPI models, {\it MPI} and {\it MPI+OpenMP},
reinforcing the trend that LLMs struggle with MPI code generation.

\subsection{Comparison with Other Models}\label{sec:results:comparison}

\begin{figure*}[th]
    \centering
    \includegraphics[width=\linewidth]{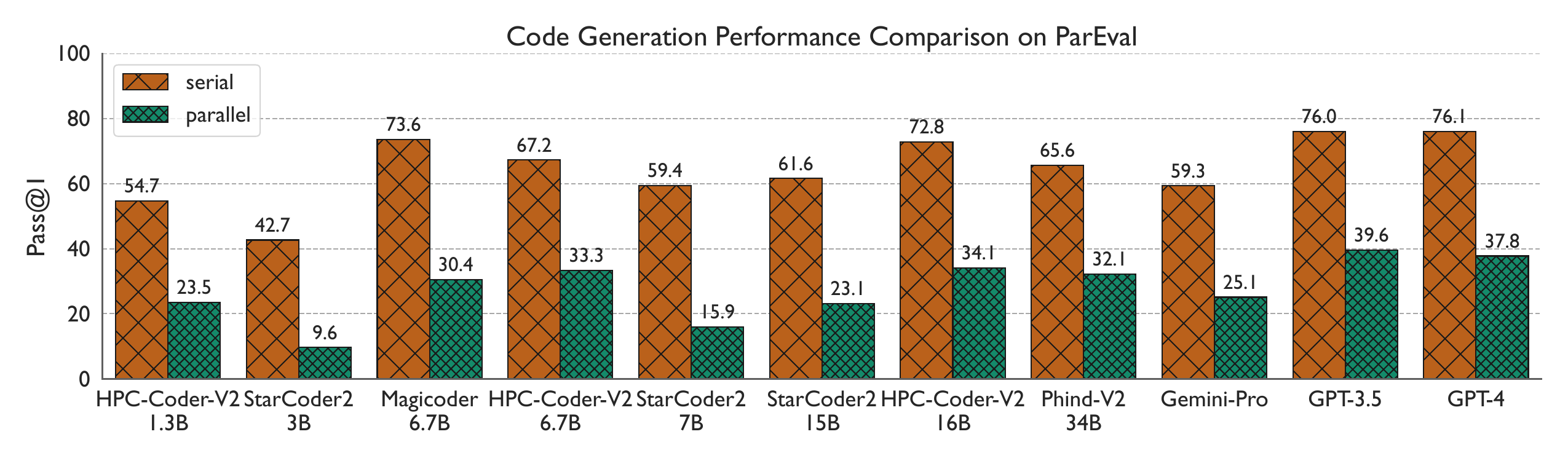}
    \caption{Comparison of ParEval parallel and serial code generation performance
        across all models. The \modelbase{} models perform as well or better than
        other models of similar size. 
    \label{fig:results:all-model-comparison}}
\end{figure*}

Finally, we compare the performance of the \modelbase{} models with other
state-of-the-art code LLMs. \Cref{fig:results:all-model-comparison} shows
ParEval parallel and serial code generation performance across all models (an
expanded list of models is shown in \Cref{sec:appendix:full-results}). We see
that, while the commercial models still dominate, the \modelbase{} models are
competitive. At each relative model size class we see that the \modelbase{}
models perform better than comparative models for parallel code generation. The
\modelbase{}-1.3B is significantly better than StarCoder2-3B despite being much
smaller. Furthermore, the \modelbase{}-6.7B model performs better than the 34B
Phind-V2 model. Despite their success at parallel code generation, the
\modelbase{} models are still beaten by Magicoder-6.7B for serial code. This
highlights, however, the success of our data and fine-tuning strategies at
training models to generate parallel code.

\begin{figure}[h]
    \centering
    \includegraphics[width=\linewidth]{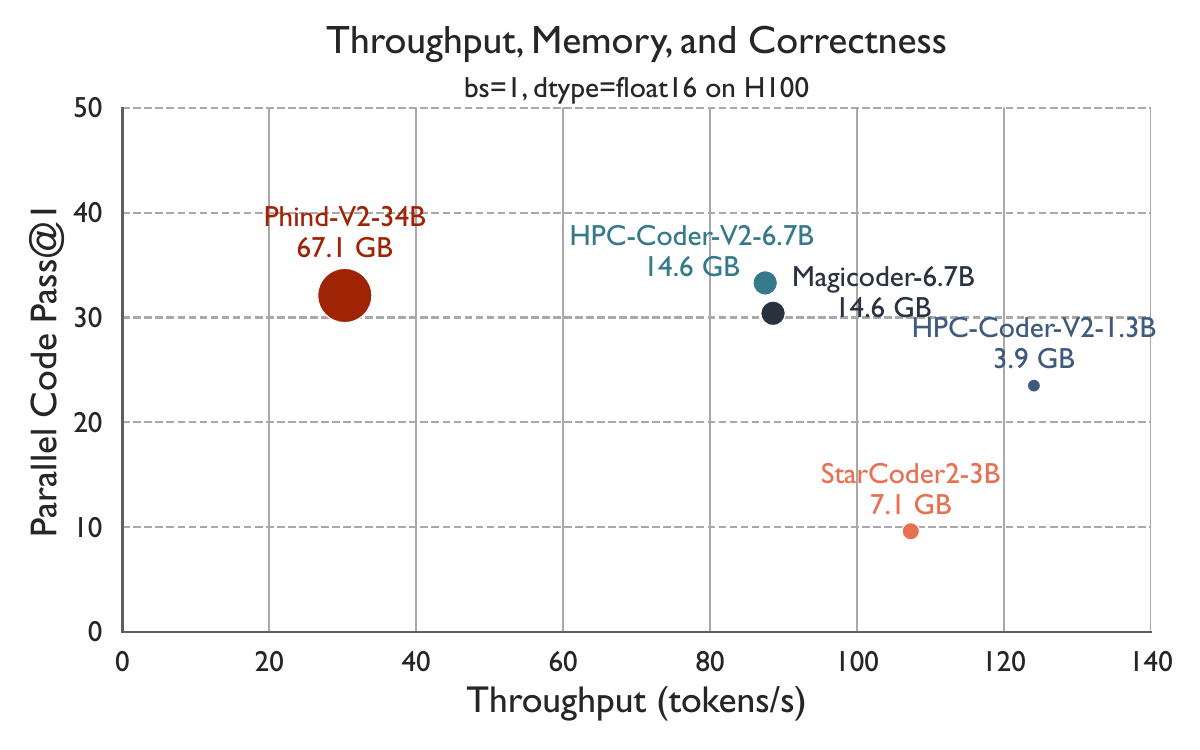}
    \caption{Comparison of parallel code generation pass rate (pass@1), model memory requirements (GB),
        and generation throughput (tokens per second). The top right of the graph is the
        ideal location where models generation correct code quickly. The smaller the dot
        the lower the model memory requirements. We see that the 6.7B model gets similar
        performance to the much larger 34B model while generating tokens significantly faster.
    \label{fig:results:throughput}}
\end{figure}

Although parallel code correctness is the most important metric for an HPC code
LLM, the system requirements of the model and the speed at which it can generate
code are also very important to developers. A model that can generate correct
code nearly as often as a larger model, but can run quickly on a consumer
laptop, is arguably much more useful for developers than the larger model. To
study this trade-off in the \modelbase{} models, we present the throughput,
required memory, and ParEval parallel pass@1 results for each model in
\Cref{fig:results:throughput}. The size of the dots are scaled based on the
memory requirement of the model with larger dots indicating larger models. The
ideal location for a model is the top right where the model generates correct
code quickly.

We see that the \modelbase{} models generate parallel code just as well or
better than the other models while being faster and more memory efficient.
\modelbase{}-6.7B is significantly faster than Phind-V2-34B while requiring much
less memory and having slightly better performance on ParEval. Magicoder-6.7B
has similar throughput and memory requirements as \modelbase{}-6.7B, but
performs worse at generating parallel code. The \modelbase{}-1.3B model is the
fastest and requires the least amount of memory, yet it outperforms other models
in its size class (StarCoder2-3B). These results demonstrate that with high
quality fine-tuning data we do not need to sacrifice memory and throughput to
generate high quality parallel code.

%% file: related-work.tex
In this section we highlight related literature on the use and design of code
LLMs for HPC and parallel code. We further discuss works focused on fine-tuning
specialized code LLMs.

\subsection{Code LLMs for HPC}\label{sec:related-work:code-llms}

Since code LLMs became popular with OpenAI's
copilot~\cite{codex-copilot-short-author} many works have focused on adapting
these models for HPC and parallel code~\cite{schneider:hpdc2024,
nichols:hpdc2024, munley2023llm4vv, chen2023data, kadosh2023scope,
nichols:isc2024, nichols2024performancealignedllmsgeneratingfast}. These works
generally fall into two categories: (1) {\it creating improved LLMs that are
better at HPC tasks} and (2) {\it engineering HPC agents and tools to leverage
existing state-of-the-art LLMs for HPC tasks}. Our work falls into the first
category, so we focus on literature in this area. However, we note that the
models and insights contributed in our work will be invaluable for studies in
the second category~\cite{yin2025chathpc, dearing:cluster2024,
munley2023llm4vv}.

Several papers that focus on creating improved LLMs for HPC tasks have focused
on more narrow tasks within HPC code generation. Schneider et
al.~\cite{schneider:hpdc2024} introduce the MPIrigen model approach for
generating MPI code. OMPGPT is introduced by Chen et al~\cite{chen2024ompgpt}
for generating OpenMP code. None of these works focus on creating general code
LLMs that can handle a wide variety of HPC tasks. The most similar to this work,
Nichols et al.~\cite{nichols:isc2024}, fine-tuned the HPC-Coder model using
scraped HPC data from GitHub and the PolyCoder base
model~\cite{xu_2022_code-llms-survey-dataset}. While this work fine-tuned a
general HPC model, the base LLM used, PolyCoder, is significantly out-of-date
compared to the state-of-the-art models used in this work. For reference,
PolyCoder is based on the GPT-2 architecture and achieves a pass@1 of 5.59\% on
the HumanEval benchmark~\cite{codex-copilot-short-author} whereas even the
smallest model used in this work,
Deepseek-Coder-1.3B~\cite{guo2024deepseekcoderlargelanguagemodel}, achieves a
pass@1 of 65.2\% on the same benchmark.

\subsection{Fine-tuning Specialized Code LLMs}\label{sec:related-work:fine-tuning}

Beyond HPC there are a great many of works that focus on fine-tuning code LLMs
for specialized tasks or domains. Tang et al~\cite{tang2023biocoder} introduce
BioCoder to address code generation tasks in the biological domain. Liu et
al~\cite{liu2023verilogeval} introduce VerilogEval for evaluating LLMs on
Verilog code generation tasks. Other works focus on more abstract issues that
arise when creating specialized code LLMs. Cassano et
al~\cite{cassano:oopsla2024} introduce a methodology for overcoming data
limitations for low-resource languages. This is aimed to aid in cases where not
enough data is available in a particular programming language to effectively
train a model. While the semi-synthetic approach in the paper may be useful for
HPC data, we found in our results that data amount was not the primary issue for
HPC code LLMs, but rather data quality. Another paper exploring both data amount
and quality by Wei et al~\cite{wei2023magicoder} uses LLM generated synthetic
data to overcome data limitations. The data collection portion of our work is an
extension of the ideas in this paper for HPC data.

%% file: conclusion.tex
In this paper we introduced a new HPC instruction dataset, \datasetname{},
using synthetic data generated from LLMs and open-source parallel code. 
Using this dataset we conduct an in-depth study along the data, model, and
prompt configuration axes of model fine-tuning to better understand how 
individual choices impact the ability of a code LLM to generate parallel code.
From this study we find the following insights:

\begin{itemize}
    \item Instruction masking during fine-tuning has little to no impact on the ability of a code LLM to generate parallel code.
    \item Fine-tuning base models, rather than their instruct variants, leads to better parallel code generation capabilities.
    \item Increasing the amount of training data for a particular parallel execution model can improve the performance of smaller code LLMs on that model with diminishing returns, but has little to no effect on larger models.
    \item The quality of the parallel code fine-tuning data can have a significant impact on the performance of a code LLM on parallel code generation.
    \item Moving from small to medium size HPC code LLMs can lead to significant improvements, while further increasing model size has diminishing returns.
\end{itemize}

Using these insights and the \datasetname{} dataset we fine-tuned three
state-of-the-art HPC code LLMs: \modelbase-1.3B, \modelbase-6.7B, and
\modelbase-16B. We evaluated these models on the ParEval benchmark and compared
them to other state-of-the-art code LLMs. We found that our models are currently
the best performing open-source models at generating parallel code. Furthermore,
our models run faster and use less memory than other models with similar or even
less parallel code generation capabilities. The models and insights contributed
in this work will be invaluable for both HPC developers and future studies into
code LLMs for HPC.

%% file: appendix.tex
\subsection{Reproducibility}\label{sec:appendix:reproducability}

The scripts used in this study are available at
\url{https://github.com/parallelcodefoundry/HPC-Coder}. The \datasetname{}
dataset can be found at
\url{https://huggingface.co/datasets/hpcgroup/hpc-instruct}. Each of the final
\modelbase{} models is available on the Hugging Face model hub at
\url{https://huggingface.co/collections/hpcgroup/hpc-coder-v2-66b681a3a8a5a1978e163a5d}.

\subsection{Example \datasetname{} Sample}\label{sec:appendix:example-sample}

The following is an example of a sample from the \datasetname{} dataset. The
problem statement and solution were generated by Llama3-70B with the seed
snippet \mintinline{cpp}{static bag_t threadbag[NUMTHREADS + 1];}.
\Cref{lst:appendix:example-problem} shows the problem statement and solution
generated by the model.

\begin{listing}[h]
    \centering
        \begin{minted}[ fontsize=\scriptsize, bgcolor=LightGray, ]{markdown} 
** Problem Statement **
In a high-performance computing application, you are 
tasked with optimizing a critical component that processes 
large datasets. The component, responsible for aggregating 
statistical metrics, is currently sequential and 
bottlenecking the entire system. Your mission is to 
parallelize the aggregation process to significantly reduce 
the processing time. The statistical metrics are stored in 
a 2D array `data` of size `ROWS x COLS`, where each element 
`data[i][j]` represents a data point. The aggregation 
function, `compute_metric`, takes a single data point as 
input and returns a metric value. The goal is to compute 
the sum of metric values for all data points in the array. 
The original sequential code is as follows: 

```c
int compute_metric(int data_point) { 
    // complex computation involving data_point 
    return result; 
} 
void aggregate_metrics(int** data, int rows, int cols) { 
    int sum = 0; 
    for (int i = 0; i < rows; i++) { 
        for (int j = 0; j < cols; j++) { 
            sum += compute_metric(data[i][j]); 
        } 
    } 
    printf("Sum of metrics: %
} 
``` 
Your task is to parallelize the `aggregate_metrics` 
function using OpenMP to take advantage of multiple CPU 
cores. The `compute_metric` function remains unchanged.
    \end{minted}
    \begin{minted}[ fontsize=\scriptsize, bgcolor=LightBlue]{markdown}
** Solution **
To parallelize the `aggregate_metrics` function, we can use 
OpenMP's parallel for directive to distribute the 
computation across multiple threads. We'll also employ 
OpenMP's reduction clause to safely accumulate the partial 
sums computed by each thread. Here's the parallelized code: 

```c 
#include <omp.h> 
int compute_metric(int data_point) { 
    // complex computation involving data_point 
    return result; 
} 

void aggregate_metrics(int** data, int rows, int cols) { 
    int sum = 0; 
    #pragma omp parallel for reduction(+:sum) 
    for (int i = 0; i < rows; i++) { 
        for (int j = 0; j < cols; j++) { 
            sum += compute_metric(data[i][j]); 
        } 
    } 
    printf("Sum of metrics: %
} 
``` 
In this solution: 

* We added the `#pragma omp parallel for` directive to 
  parallelize the outer loop, which iterates over the rows 
  of the 2D array. 
* We used the `reduction(+:sum)` clause to specify that 
  each thread should maintain its own partial sum, which 
  will be combined using the `+` operator at the end of the 
  parallel region. This ensures that the final sum is 
  correctly computed. 
* The inner loop, which iterates over the columns, is 
  executed sequentially within each thread, as it has no 
  dependencies between iterations. By parallelizing the 
  `aggregate_metrics` function, we can significantly reduce 
  the processing time for large datasets, taking advantage 
  of the available CPU cores.
    \end{minted}
    \caption{Example problem statement from the \datasetname{} dataset.
    The problem requires the model to parallelize a for-loop using OpenMP.
    \label{lst:appendix:example-problem}}
\end{listing}

\subsection{Full ParEval Results}\label{sec:appendix:full-results}

\begin{figure*}[h]
    \centering
    \begin{tabular}{cc}
        \subcaptionbox{\modelbase-1.3B\label{fig:results:pass1-heatmap-1.3B}}{\includegraphics[width=0.5\linewidth]{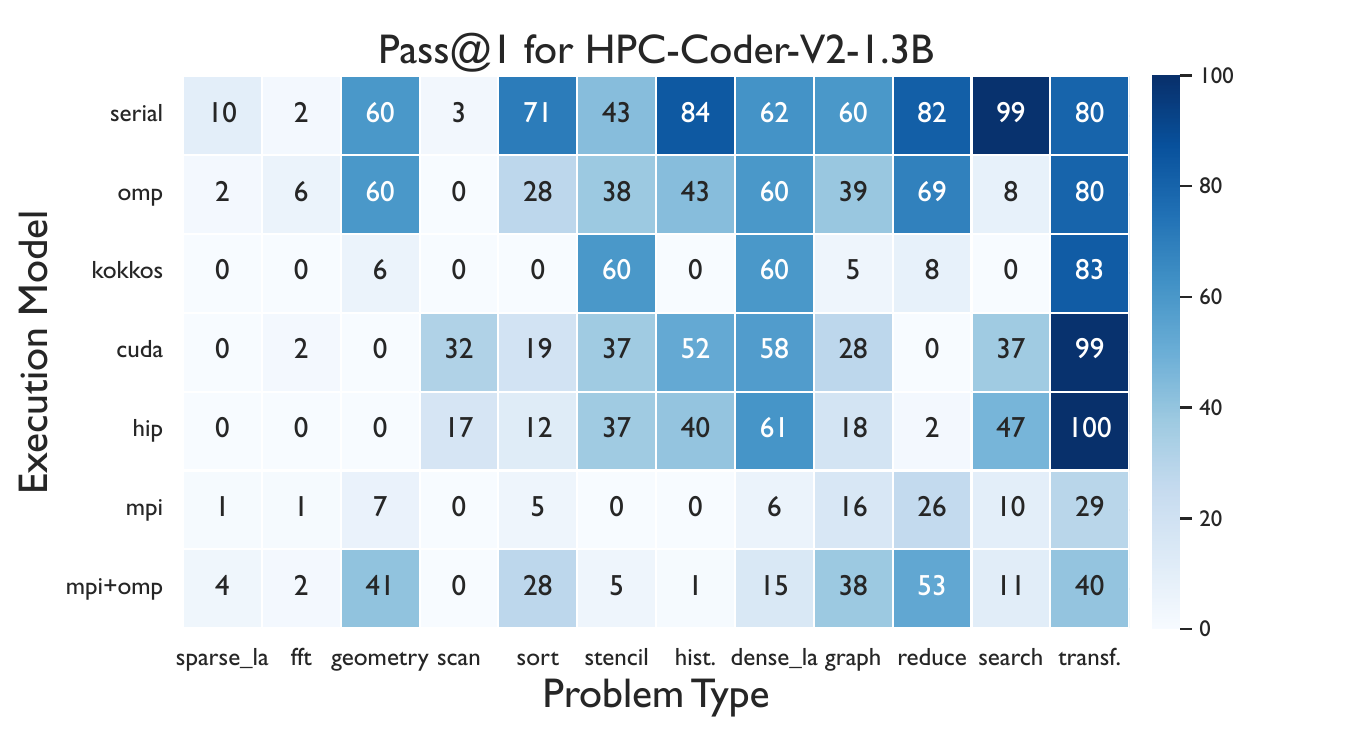}} &
        \subcaptionbox{StarCoder2-3B\label{fig:results:pass1-heatmap-starcoder2-3B}}{\includegraphics[width=0.5\linewidth]{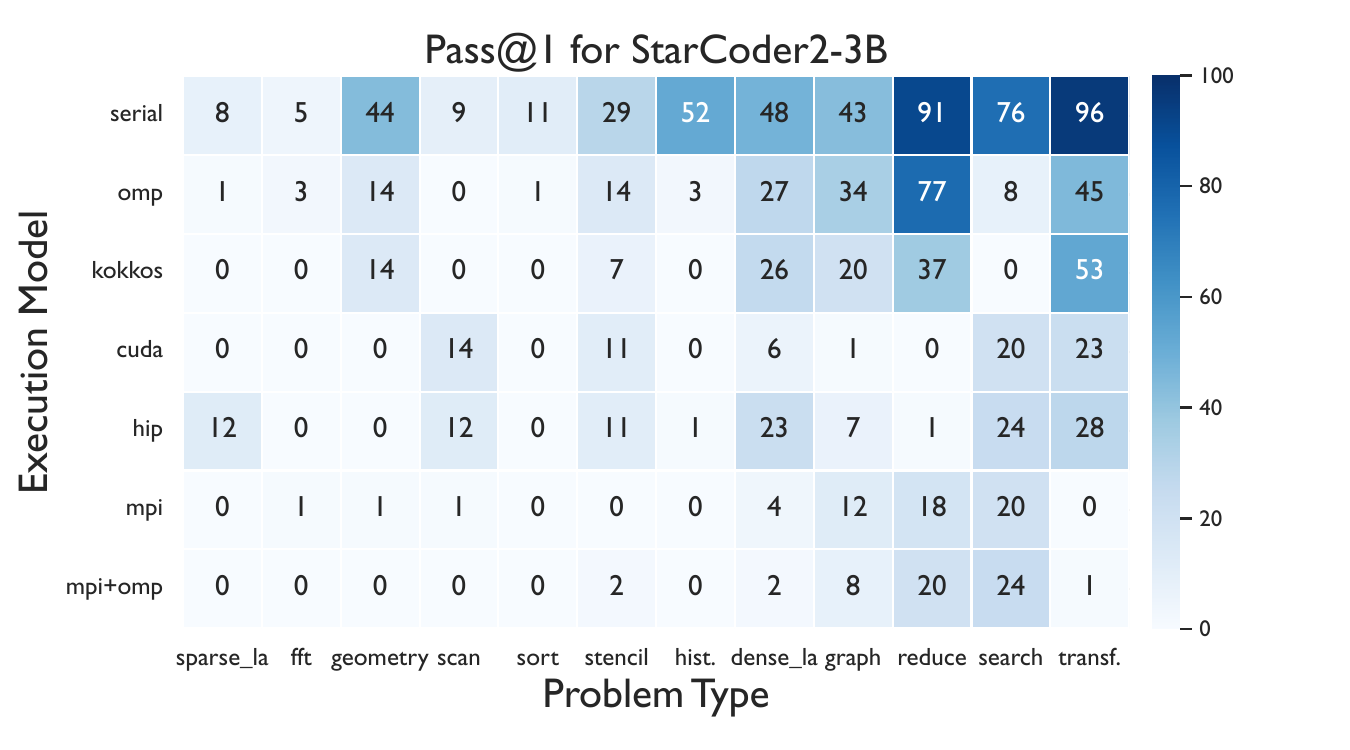}} \\
        \subcaptionbox{\modelbase-6.7B\label{fig:results:pass1-heatmap-6.7B}}{\includegraphics[width=0.5\linewidth]{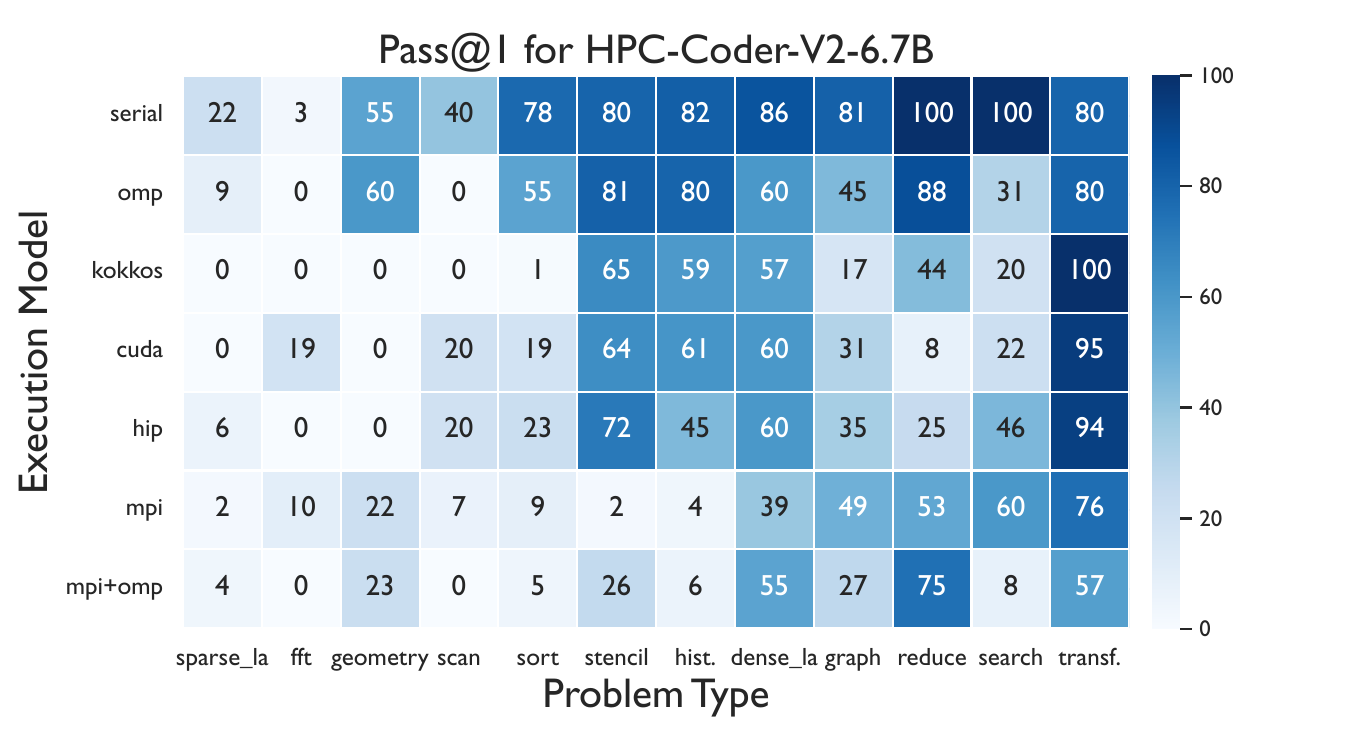}} &
        \subcaptionbox{\modelbase-16B\label{fig:results:pass1-heatmap-16B}}{\includegraphics[width=0.5\linewidth]{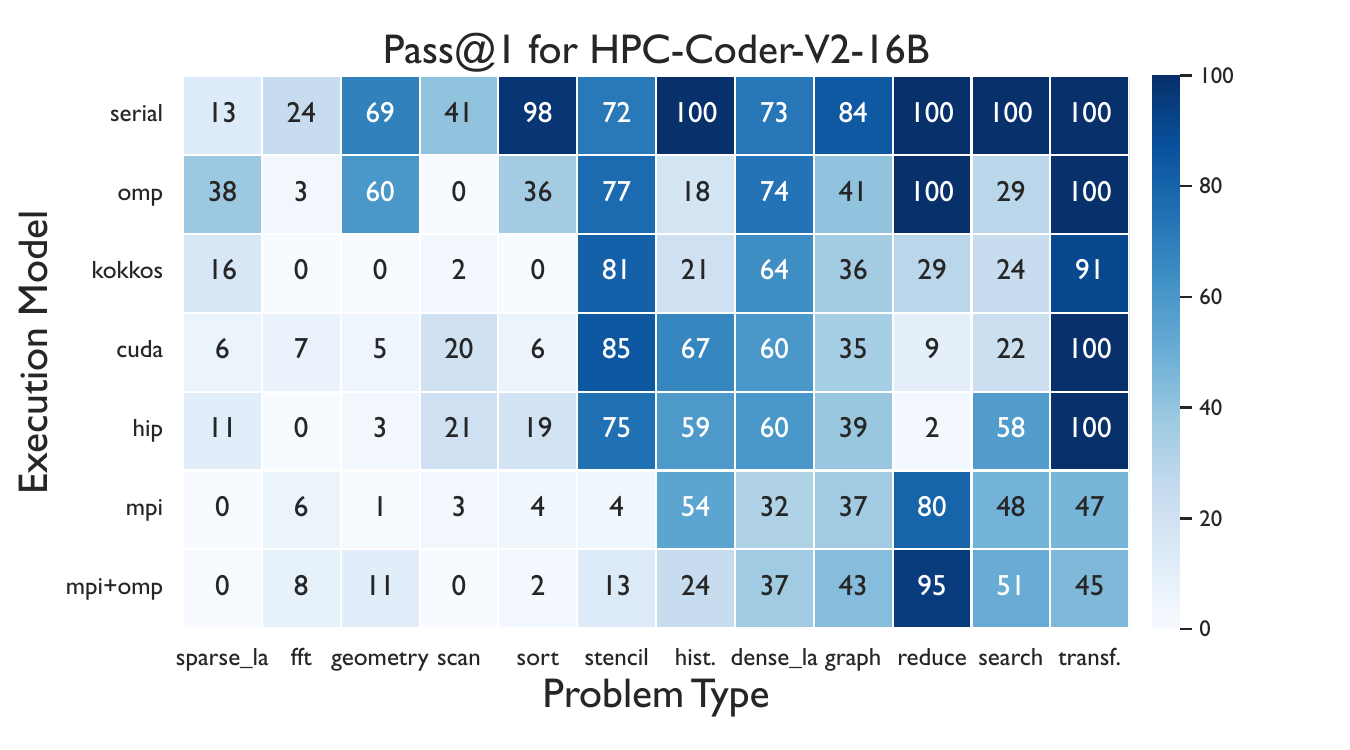}} \\
        \subcaptionbox{Phind-V2-34B\label{fig:results:pass1-heatmap-Phind-V2-34B}}{\includegraphics[width=0.5\linewidth]{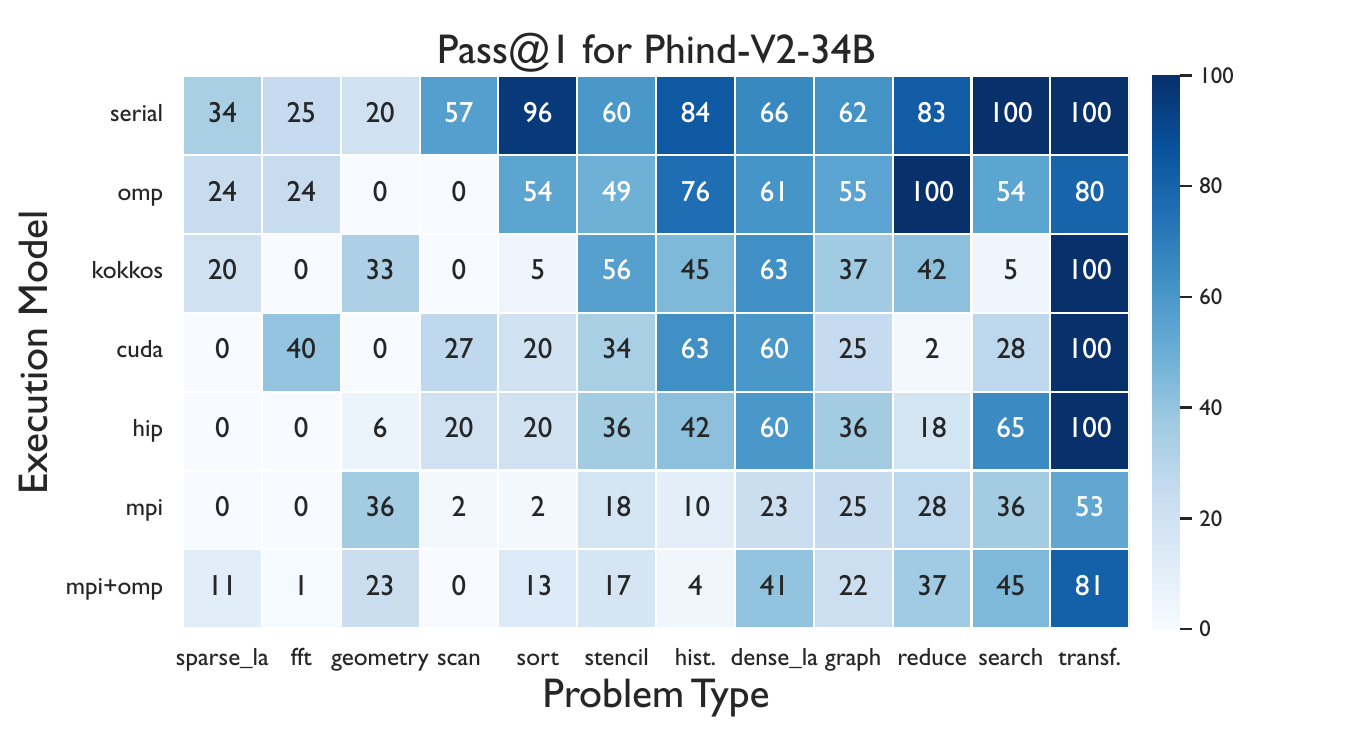}} &
        \subcaptionbox{GPT-4\label{fig:results:pass1-heatmap-GPT-4}}{\includegraphics[width=0.5\linewidth]{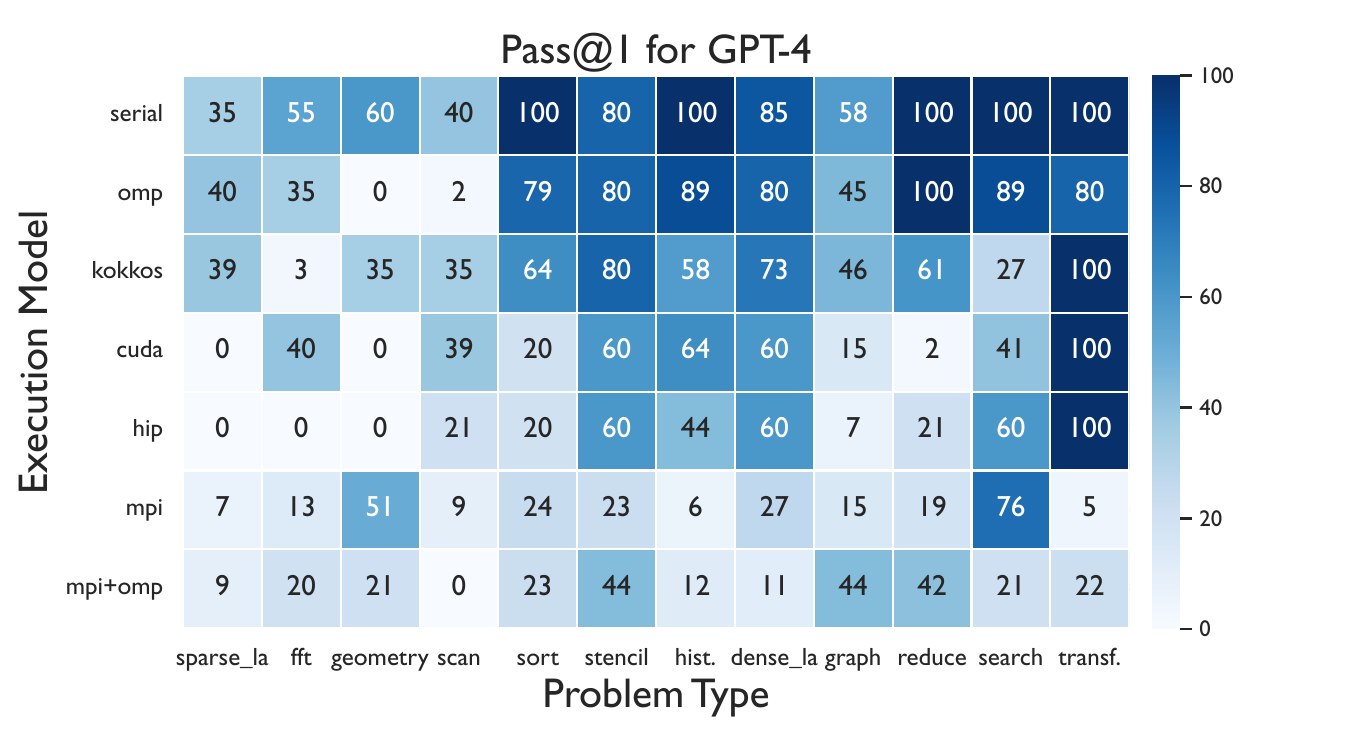}} \\
    \end{tabular}
    \caption{Complete generation results for a sample of the models on the ParEval benchmark.
    Each box shows the pass@1 score for a problem type and parallel execution model.
    \label{fig:results:pass1-heatmap}}
\end{figure*}

Complete ParEval correctness results for all \modelbase{} models, StarCoder2-3B,
Phind-V2-34B, and GPT-4 are shown in Figure~\ref{fig:results:pass1-heatmap}. Each
box in the heatmap represents the estimated pass@1 for the corresponding model
on a particular problem type and execution model. The number is estimated using 
twenty generated samples for each of the five problems per problem type (100 
generations total per box). Summary pass@1 results for even more models are
shown in~\Cref{tab:appendix:all-results}.

\begin{table}[h]
    \centering
    \caption{ParEval code generation results for all models.\label{tab:appendix:all-results}}
    \input{figs/all-pareval-results-table}
\end{table}

%% file: figs/all-pareval-results-table.tex
\begin{tabular}{lrrr}
\toprule
Model & Size (B) & \multicolumn{2}{c}{Pass@1} \\
 &  & serial & parallel \\
\midrule
HPC-Coder-V2-1.3B & 1.3 & 54.7 & 23.5 \\
StarCoder2-3B &  3 & 42.7 & 9.61 \\
HPC-Coder-V2-6.7B & 6.7 & 67.2 & 33.3 \\
Magicoder-6.7B & 6.7 & 73.6 & 30.4 \\
StarCoder2-7B &  7 & 59.4 & 15.9 \\
CodeLlama-7B &  7 & 48.4 & 15.3 \\
CodeLlama-13B & 13 & 52.8 & 17.4 \\
StarCoder2-15B & 15 & 61.6 & 23.1 \\
StarCoderBase & 15.5 & 51.7 & 18.6 \\
HPC-Coder-V2-16B & 16 & 72.8 & 34.1 \\
Phind-V2-34B & 34 & 65.6 & 32.1 \\
CodeLlama-34B & 34 & 54 & 10.2 \\
Gemini-Pro & --- & 59.3 & 25.1 \\
GPT-3.5 & 175 & 76 & 39.6 \\
GPT-4 & --- & 76.1 & 37.8 \\
\bottomrule
\end{tabular}

%% file: paper.bbl
\begin{thebibliography}{10}
\providecommand{\url}[1]{#1}
\csname url@samestyle\endcsname
\providecommand{\newblock}{\relax}
\providecommand{\bibinfo}[2]{#2}
\providecommand{\BIBentrySTDinterwordspacing}{\spaceskip=0pt\relax}
\providecommand{\BIBentryALTinterwordstretchfactor}{4}
\providecommand{\BIBentryALTinterwordspacing}{\spaceskip=\fontdimen2\font plus
\BIBentryALTinterwordstretchfactor\fontdimen3\font minus
  \fontdimen4\font\relax}
\providecommand{\BIBforeignlanguage}[2]{{%
\expandafter\ifx\csname l@#1\endcsname\relax
\typeout{** WARNING: IEEEtran.bst: No hyphenation pattern has been}%
\typeout{** loaded for the language `#1'. Using the pattern for}%
\typeout{** the default language instead.}%
\else
\language=\csname l@#1\endcsname
\fi
#2}}
\providecommand{\BIBdecl}{\relax}
\BIBdecl

\bibitem{github-ai-survey}
I.~Shani, ``Survey reveals ai’s impact on the developer experience,''
  \url{https://github.blog/news-insights/research/survey-reveals-ais-impact-on-the-developer-experience/},
  June 2023, accessed: 2024-10-12.

\bibitem{nichols:hpdc2024}
D.~Nichols, J.~H. Davis, Z.~Xie, A.~Rajaram, and A.~Bhatele, ``Can large
  language models write parallel code?'' in \emph{Proceedings of the 33rd
  International Symposium on High-Performance Parallel and Distributed
  Computing}, ser. HPDC '24.\hskip 1em plus 0.5em minus 0.4em\relax New York,
  NY, USA: Association for Computing Machinery, 2024.

\bibitem{lozhkov2024starcoder}
A.~Lozhkov, R.~Li, L.~B. Allal, F.~Cassano, J.~Lamy-Poirier, N.~Tazi, A.~Tang,
  D.~Pykhtar, J.~Liu, Y.~Wei, T.~Liu, M.~Tian, D.~Kocetkov, A.~Zucker,
  Y.~Belkada, Z.~Wang, Q.~Liu, D.~Abulkhanov, I.~Paul, Z.~Li, W.-D. Li,
  M.~Risdal, J.~Li, J.~Zhu, T.~Y. Zhuo, E.~Zheltonozhskii, N.~O.~O. Dade,
  W.~Yu, L.~Krauß, N.~Jain, Y.~Su, X.~He, M.~Dey, E.~Abati, Y.~Chai,
  N.~Muennighoff, X.~Tang, M.~Oblokulov, C.~Akiki, M.~Marone, C.~Mou,
  M.~Mishra, A.~Gu, B.~Hui, T.~Dao, A.~Zebaze, O.~Dehaene, N.~Patry, C.~Xu,
  J.~McAuley, H.~Hu, T.~Scholak, S.~Paquet, J.~Robinson, C.~J. Anderson,
  N.~Chapados, M.~Patwary, N.~Tajbakhsh, Y.~Jernite, C.~M. Ferrandis, L.~Zhang,
  S.~Hughes, T.~Wolf, A.~Guha, L.~von Werra, and H.~de~Vries, ``Starcoder 2 and
  the stack v2: The next generation,'' 2024.

\bibitem{vaswani2017attention}
A.~Vaswani, N.~Shazeer, N.~Parmar, J.~Uszkoreit, L.~Jones, A.~N. Gomez,
  {\L}.~Kaiser, and I.~Polosukhin, ``Attention is all you need,'' in
  \emph{Advances in neural information processing systems}, 2017, pp.
  5998--6008.

\bibitem{roziere2023code}
B.~Rozière, J.~Gehring, F.~Gloeckle, S.~Sootla, I.~Gat, X.~E. Tan, Y.~Adi,
  J.~Liu, T.~Remez, J.~Rapin, A.~Kozhevnikov, I.~Evtimov, J.~Bitton, M.~Bhatt,
  C.~C. Ferrer, A.~Grattafiori, W.~Xiong, A.~Défossez, J.~Copet, F.~Azhar,
  H.~Touvron, L.~Martin, N.~Usunier, T.~Scialom, and G.~Synnaeve, ``Code llama:
  Open foundation models for code,'' 2023.

\bibitem{openai2024gpt4ocard_short_author}
\BIBentryALTinterwordspacing
OpenAI, A.~Hurst, and et~al, ``Gpt-4o system card,'' 2024. [Online]. Available:
  \url{https://arxiv.org/abs/2410.21276}
\BIBentrySTDinterwordspacing

\bibitem{xu2024surveyknowledgedistillationlarge}
\BIBentryALTinterwordspacing
X.~Xu, M.~Li, C.~Tao, T.~Shen, R.~Cheng, J.~Li, C.~Xu, D.~Tao, and T.~Zhou, ``A
  survey on knowledge distillation of large language models,'' 2024. [Online].
  Available: \url{https://arxiv.org/abs/2402.13116}
\BIBentrySTDinterwordspacing

\bibitem{wei2023magicoder}
Y.~Wei, Z.~Wang, J.~Liu, Y.~Ding, and L.~Zhang, ``Magicoder: Source code is all
  you need,'' \emph{arXiv preprint arXiv:2312.02120}, 2023.

\bibitem{bigcode_leaderboard}
\BIBentryALTinterwordspacing
``Big code models leaderboard - a hugging face space by bigcode,'' 2023.
  [Online]. Available:
  \url{https://huggingface.co/spaces/bigcode/bigcode-models-leaderboard}
\BIBentrySTDinterwordspacing

\bibitem{guo2024deepseekcoder}
D.~Guo, Q.~Zhu, D.~Yang, Z.~Xie, K.~Dong, W.~Zhang, G.~Chen, X.~Bi, Y.~Wu,
  Y.~K. Li, F.~Luo, Y.~Xiong, and W.~Liang, ``Deepseek-coder: When the large
  language model meets programming -- the rise of code intelligence,'' 2024.

\bibitem{deepseek-coder-v2}
\BIBentryALTinterwordspacing
DeepSeek-AI, Q.~Zhu, D.~Guo, Z.~Shao, D.~Yang, P.~Wang, R.~Xu, Y.~Wu, Y.~Li,
  H.~Gao, S.~Ma, W.~Zeng, X.~Bi, Z.~Gu, H.~Xu, D.~Dai, K.~Dong, L.~Zhang,
  Y.~Piao, Z.~Gou, Z.~Xie, Z.~Hao, B.~Wang, J.~Song, D.~Chen, X.~Xie, K.~Guan,
  Y.~You, A.~Liu, Q.~Du, W.~Gao, X.~Lu, Q.~Chen, Y.~Wang, C.~Deng, J.~Li,
  C.~Zhao, C.~Ruan, F.~Luo, and W.~Liang, ``Deepseek-coder-v2: Breaking the
  barrier of closed-source models in code intelligence,'' 2024. [Online].
  Available: \url{https://arxiv.org/abs/2406.11931}
\BIBentrySTDinterwordspacing

\bibitem{touvron2023llama}
H.~Touvron \emph{et~al.}, ``Llama 2: Open foundation and fine-tuned chat
  models,'' Tech. Rep., 2023.

\bibitem{fb-moe}
\BIBentryALTinterwordspacing
M.~Artetxe, S.~Bhosale, N.~Goyal, T.~Mihaylov, M.~Ott, S.~Shleifer, X.~V. Lin,
  J.~Du, S.~Iyer, R.~Pasunuru, G.~Anantharaman, X.~Li, S.~Chen, H.~Akin,
  M.~Baines, L.~Martin, X.~Zhou, P.~S. Koura, B.~O'Horo, J.~Wang,
  L.~Zettlemoyer, M.~Diab, Z.~Kozareva, and V.~Stoyanov, ``Efficient large
  scale language modeling with mixtures of experts,'' 2021. [Online].
  Available: \url{https://arxiv.org/abs/2112.10684}
\BIBentrySTDinterwordspacing

\bibitem{luo2023wizardcoder}
Z.~Luo, C.~Xu, P.~Zhao, Q.~Sun, X.~Geng, W.~Hu, C.~Tao, J.~Ma, Q.~Lin, and
  D.~Jiang, ``Wizardcoder: Empowering code large language models with
  evol-instruct,'' \emph{arXiv preprint arXiv:2306.08568}, 2023.

\bibitem{singh:ipdps2023}
\BIBentryALTinterwordspacing
S.~Singh and A.~Bhatele, ``Exploiting sparsity in pruned neural networks to
  optimize large model training,'' in \emph{2023 IEEE International Parallel
  and Distributed Processing Symposium (IPDPS)}.\hskip 1em plus 0.5em minus
  0.4em\relax Los Alamitos, CA, USA: IEEE Computer Society, may 2023, pp.
  245--255. [Online]. Available:
  \url{https://doi.ieeecomputersociety.org/10.1109/IPDPS54959.2023.00033}
\BIBentrySTDinterwordspacing

\bibitem{paszke2019pytorch}
A.~Paszke, S.~Gross, F.~Massa, A.~Lerer, J.~Bradbury, G.~Chanan, T.~Killeen,
  Z.~Lin, N.~Gimelshein, L.~Antiga, A.~Desmaison, A.~Köpf, E.~Yang, Z.~DeVito,
  M.~Raison, A.~Tejani, S.~Chilamkurthy, B.~Steiner, L.~Fang, J.~Bai, and
  S.~Chintala, ``Pytorch: An imperative style, high-performance deep learning
  library,'' 2019.

\bibitem{adamw}
\BIBentryALTinterwordspacing
I.~Loshchilov and F.~Hutter, ``Fixing weight decay regularization in adam,''
  \emph{CoRR}, vol. abs/1711.05101, 2017. [Online]. Available:
  \url{http://arxiv.org/abs/1711.05101}
\BIBentrySTDinterwordspacing

\bibitem{codex-copilot-short-author}
M.~Chen and et~al, ``Evaluating large language models trained on code,'' 2021.

\bibitem{phind-codellama-34b-v2}
\BIBentryALTinterwordspacing
Phind. (2023) Phind-codellama-34b-v2. [Online]. Available:
  \url{https://huggingface.co/Phind/Phind-CodeLlama-34B-v2}
\BIBentrySTDinterwordspacing

\bibitem{geminishort2023gemini}
G.~Team, ``Gemini: A family of highly capable multimodal models,'' 2023.

\bibitem{gpt-3}
\BIBentryALTinterwordspacing
T.~B. Brown \emph{et~al.}, ``Language models are few-shot learners,''
  \emph{CoRR}, vol. abs/2005.14165, 2020. [Online]. Available:
  \url{https://arxiv.org/abs/2005.14165}
\BIBentrySTDinterwordspacing

\bibitem{openai2023gpt4}
OpenAI, ``Gpt-4 technical report,'' 2023.

\bibitem{schneider:hpdc2024}
\BIBentryALTinterwordspacing
N.~Schneider, N.~Hasabnis, V.~A. Vo, T.~Kadosh, N.~Krien, M.~Capota, G.~Tamir,
  T.~L. Willke, N.~Ahmed, Y.~Pinter, T.~Mattson, and G.~Oren, ``Mpirigen: Mpi
  code generation through domain-specific language models,'' in
  \emph{Proceedings of the 2024 Workshop on AI For Systems}, ser. AI4Sys
  '24.\hskip 1em plus 0.5em minus 0.4em\relax New York, NY, USA: Association
  for Computing Machinery, 2024, p. 1–6. [Online]. Available:
  \url{https://doi.org/10.1145/3660605.3660944}
\BIBentrySTDinterwordspacing

\bibitem{munley2023llm4vv}
C.~Munley, A.~Jarmusch, and S.~Chandrasekaran, ``Llm4vv: Developing llm-driven
  testsuite for compiler validation,'' 2023.

\bibitem{chen2023data}
L.~Chen, X.~Ding, M.~Emani, T.~Vanderbruggen, P.~hung Lin, and C.~Liao, ``Data
  race detection using large language models,'' 2023.

\bibitem{kadosh2023scope}
T.~Kadosh, N.~Hasabnis, V.~A. Vo, N.~Schneider, N.~Krien, A.~Wasay, N.~Ahmed,
  T.~Willke, G.~Tamir, Y.~Pinter, T.~Mattson, and G.~Oren, ``Scope is all you
  need: Transforming llms for hpc code,'' 2023.

\bibitem{nichols:isc2024}
D.~Nichols, A.~Marathe, H.~Menon, T.~Gamblin, and A.~Bhatele, ``Modeling
  parallel programs using large language models,'' ser. ISC '24, may 2024.

\bibitem{nichols2024performancealignedllmsgeneratingfast}
\BIBentryALTinterwordspacing
D.~Nichols, P.~Polasam, H.~Menon, A.~Marathe, T.~Gamblin, and A.~Bhatele,
  ``Performance-aligned llms for generating fast code,'' 2024. [Online].
  Available: \url{https://arxiv.org/abs/2404.18864}
\BIBentrySTDinterwordspacing

\bibitem{yin2025chathpc}
J.~Yin, J.~Hines, E.~Herron, T.~Ghosal, H.~Liu, S.~Prentice, V.~Lama, and
  F.~Wang, ``chathpc: Empowering hpc users with large language models,''
  \emph{The Journal of Supercomputing}, vol.~81, no.~1, p. 194, 2025.

\bibitem{dearing:cluster2024}
M.~T. Dearing, Y.~Tao, X.~Wu, Z.~Lan, and V.~Taylor, ``Lassi: An llm-based
  automated self-correcting pipeline for translating parallel scientific
  codes,'' in \emph{2024 IEEE International Conference on Cluster Computing
  Workshops (CLUSTER Workshops)}, 2024, pp. 136--143.

\bibitem{chen2024ompgpt}
L.~Chen, A.~Bhattacharjee, N.~Ahmed, N.~Hasabnis, G.~Oren, V.~Vo, and
  A.~Jannesari, ``Ompgpt: A generative pre-trained transformer model for
  openmp,'' \emph{arXiv preprint arXiv:2401.16445}, 2024.

\bibitem{xu_2022_code-llms-survey-dataset}
\BIBentryALTinterwordspacing
F.~F. Xu, U.~Alon, G.~Neubig, and V.~J. Hellendoorn, ``{A Systematic Evaluation
  of Large Language Models of Code},'' Feb. 2022,
  https://arxiv.org/abs/2202.13169. [Online]. Available:
  \url{https://doi.org/10.5281/zenodo.6363556}
\BIBentrySTDinterwordspacing

\bibitem{guo2024deepseekcoderlargelanguagemodel}
\BIBentryALTinterwordspacing
D.~Guo, Q.~Zhu, D.~Yang, Z.~Xie, K.~Dong, W.~Zhang, G.~Chen, X.~Bi, Y.~Wu,
  Y.~K. Li, F.~Luo, Y.~Xiong, and W.~Liang, ``Deepseek-coder: When the large
  language model meets programming -- the rise of code intelligence,'' 2024.
  [Online]. Available: \url{https://arxiv.org/abs/2401.14196}
\BIBentrySTDinterwordspacing

\bibitem{tang2023biocoder}
X.~Tang, B.~Qian, R.~Gao, J.~Chen, X.~Chen, and M.~Gerstein, ``Biocoder: A
  benchmark for bioinformatics code generation with contextual pragmatic
  knowledge,'' 2023.

\bibitem{liu2023verilogeval}
M.~Liu, N.~Pinckney, B.~Khailany, and H.~Ren, ``Verilogeval: Evaluating large
  language models for verilog code generation,'' 2023.

\bibitem{cassano:oopsla2024}
\BIBentryALTinterwordspacing
F.~Cassano, J.~Gouwar, F.~Lucchetti, C.~Schlesinger, A.~Freeman, C.~J.
  Anderson, M.~Q. Feldman, M.~Greenberg, A.~Jangda, and A.~Guha, ``Knowledge
  transfer from high-resource to low-resource programming languages for code
  llms,'' \emph{Proc. ACM Program. Lang.}, vol.~8, no. OOPSLA2, Oct. 2024.
  [Online]. Available: \url{https://doi.org/10.1145/3689735}
\BIBentrySTDinterwordspacing

\end{thebibliography}
